\documentclass[aip,  jcp,  reprint]{revtex4-2}
\usepackage{amsmath}
\usepackage{amssymb}
\usepackage{cases}
\usepackage{upgreek}
\usepackage{graphicx}
\usepackage{xcolor}
\begin{document}

\title{Thermodynamics of a compressible lattice gas  crystal: 
Generalized Gibbs-Duhem equation and adsorption}

\author{Michiel Sprik}
\email{ms284@cam.ac.uk}
\affiliation{Yusuf Hamied Department of Chemistry, University of Cambridge, Lensfield Road, Cambridge CB2 1EW, United Kingdom}

\begin{abstract}
Compressible lattice gas models are used in material science to understand the coupling between composition and strain in alloys. The seminal work in this field is the 1973 Larch\'{e}-Cahn paper (Acta Metall. {\bf 21} 1051-1063). Single-phase crystals in Larch\'{e}-Cahn theory are stable under open constant pressure, constant temperature conditions. The Gibbs free energy does not have to match the product $\mu N$ of the number of particles $N$ and their chemical potential $\mu$.  Similarly, the grand potential and the product $pV$ of pressure and volume $V$ may not add up to zero. Discrepancies already arise under hydrostatic stress. The elastic energy is not proportional to volume and the Gibbs-Duhem relation valid for liquids is violated. Extensivity is recovered by treating the number of lattice sites $M$ as an additional thermodynamic variable. The difference $ G-\mu N $ can be identified with $\nu M$ where $\nu$ is the thermodynamic force conjugate to $M$. The reinstated Gibbs-Duhem equation can be cast in the form of an adsorption equation and applied to quantify the tendency to vacancy creation under isothermal isobaric conditions.  We have worked this out for a uniform one-component compressible lattice gas crystal. Shear stress is omitted. The coupling between composition and strain is implemented by decomposing  pressure in a mechanical component depending on deformed density $N/V$ and an elastic term linear in the volume strain as determined by $V/M$.  Various $\left( \mu, p, T \right) $ response functions are compared to the $\left( \mu, V, T \right) $ counterparts.
\end{abstract}

\maketitle

\section{Introduction} \label{sec:intro}

What is the difference between the macroscopic thermodynamics of a homogeneous solid and a liquid? This question already raised by Gibbs is of the utmost importance  in material science\cite{gibbscw, cahn85}.  It cannot only be the shear elasticity.  Solids exhibit defects which are non-existent or relax away in liquids\cite{cahn85}.  In particular,  point defects,  such as vacancies and interstitials,  are structural features of equilibrium solids.  Vacancies and diffusion in solids were not known at the time Gibbs created his theory of chemical thermodynamics. This caused him to question whether it is possible to define a proper chemical potential for solids similar to the chemical potential for liquids.  While in principle atoms can be removed  from a solid, inserting one is virtually impossible.  Atoms can only be added at the surface in the view of Gibbs,  a process that is usually referred to as accretion.  Coupling to a chemical potential of the solid is not required for the thermodynamics of accretion.  

After experiment had established the existence of vacancies and shown that they are also mobile,  the concept of a chemical potential for solids was introduced by the founding fathers of modern material science\cite{cahn73,cahn78, mullins84,mullins85,sekerka89} (for reviews see Refs.~\citenum{cahn85,larche96} and \citenum{voorhees04}).  These theories are intimately related to the question of the fundamental distinction between a solid and liquid.  This was formulated in terms of a lattice or network that supports diffusion but is not affected by it.  Nodes of this network can only be added or destroyed at interfaces with liquids or internal surface and line defects,  such as grain boundaries and dislocations. This is the general consensus in material science but details can differ.  The approach that has become the basis for understanding of  coupling between composition and strain is the network theory of Larch\'{e} and Cahn\cite{cahn73,cahn78}.  In their view atoms can only move by exchanging lattice site with a vacancy.  This restricts the definition of chemical potential to relative values,  called diffusion potentials (for a somewhat different view Mullins and Sekerka\cite{mullins84,mullins85,sekerka89}).  

The Larch\'{e}-Cahn (LC) theory has been applied  to explain a wide range of problems involving metal alloys\cite{weissmueller18}  and also ceramic composites.  However there  still remain issues causing confusion.  One such issue is the question of the validity of the Gibbs-Duhem equation.  A reason to doubt this is that elastic energy functions are not manifest extensive (as noted for example in Refs.  \citenum{gurtin99, gurtin10}).  Strain is defined as a deformation relative to a fixed reference system. Shear deformation changes the shape of a crystal preserving volume. However, isotropic dilation and contraction directly changes the size.  The corresponding elastic energy density in the small deformation approximation scales with the square of volume violating  extensivity as required in the usual derivation of the Gibbs-Duhem relation.  This also must have repercussions for a consistent definition of chemical potential of an elastic solid. 

Extensivity of elastic energy can be restored by including reference volume in the thermodynamics degrees of freedom of the solid. Homogeneous increase in system size can then be interpreted as multiplying volume and reference volume by the same scaling factor.  The volume stretch, from which the strain is derived,  is the ratio of volume and reference volume and is invariant under this generalized scaling operation.  The same applies to the elastic energy density. As a result  the total energy is again proportional to system size satisfying the conditions or a generalized Gibbs-Duhem relation. However reference volume has not quite the same thermodynamic status as the actual (deformed) volume. Reference volume is an extensive order parameter for solid rigidity.  It plays no role in liquids or stated more formally, the free energy of a liquid should be invariant under a change of reference volume. 

LC theory is based on a lattice model. The number of lattice sites is invariant under elastic deformation. This is a fundamental postulate of the LC network concept\cite{cahn73}. The cell volume of the reference lattice  is constant and therefore the reference volume for the deformable lattice system is determined by the number of lattice sites denoted by $M$ (see also Ref.~\citenum{mullins85}). $M$ is the configurational degree of freedom representing  reference volume in LC theory.   Moreover, $M$ is conserved and we are therefore allowed to declare $M$ to be an extensive variable in addition to the number of particles $N$ and deformed volume $V$. The thermodynamic forces conjugate to $N$ and $V$ are the chemical potential $\mu$  respectively (minus) the pressure  $p$. The thermodynamic field conjugate to $M$ will be denoted by $\nu$. The question is now what is this $\nu$ and how can it be used in thermodynamic derivations. This is the main topic of the present paper. 

The  author failed to appreciate until recently the significance of the LC network principle and the special role of volume elasticity.   In this respect work written up in two previous papers on solid thermodynamics  is fundamentally flawed\cite{sprik21c,sprik24}.  Hopefully this will be redressed by the  present study on a single component compressible lattice gas model.  Chemomechanical coupling is accounted for by separating  pressure  in a molecular component determined by the deformed density $N/V$ and an elastic component depending on deformation as quantified by deformed cell volume $V/M$.     While easily set up in a  discrete lattice model, from a statistical mechanical perspective this decomposition of the pressure requires justification in terms of structural correlations. We are not yet able   to provide such a microscopic underpinning.  The focus of the paper is on the implications of this mechanism of chemomechanical  coupling under hydrostatic pressure.  Shear elasticity is ignored. 

Staying within the framework of LC theory, we derive an expression for the thermodynamic force conjugate to $M$. This quantity,  which we gave the symbol $\nu$, is discussed in some detail in the Voorhees-Johnson  review\cite{voorhees04}. Here we will put $\nu$ to work.  As will be shown,  the differential of $\nu$ defines a Gibbs absorption isotherm which can be used to evaluate the isothermal thermodynamic response coefficients of the LC model under open isobaric isothermal conditions $\left( \mu, P, T \right)$.  The $\left( \mu, P, T \right)$ response coefficients are compared to results for the  $\left( \mu, V, T \right)$ system obtained by direct differentiation. The main conclusion is that differences are more prominent in soft systems as quantified by the closed system compressibility.  

The structure of the paper is as follows. The generalized $\left(N,V,M\right)$ thermodynamics is reviewed in section \ref{sec:thermech}. Section \ref{sec:rigid} is a brief recapitulation of the Langmuir lattice gas model for a rigid solid introducing the number of lattice sites $M$ as a thermodynamic variable in addition to number of particles $N$. The chemical potential $\mu$ and lattice potential $\nu$ are derived and it is shown how the differential of $\nu$  can be interpreted as a Gibbs adsorption isotherm.  In section \ref{sec:model} the rigid lattice model is generalized to a model for a compressible lattice.   In section \ref{sec:dtherm},  following the scheme of section \ref{sec:rigid},  the expression for the now strain dependent thermodynamic forces  $\mu$ and $\nu$ are derived  by partial differentiation with respect to $N$ respectively $M$.  The  pressure is obtained from the corresponding volume derivative.  In section \ref{sec:epstrain} the  model for occupation-strain coupling introduced in section \ref{sec:model} is expanded to second order in occupation and checked for consistence by verifying that the conditions for accretion are verified in this approximation.   In section \ref{sec:response} the theory is applied deriving the expression for various thermodynamic response coefficients.   As an application we work out the equation for the change in density in response to open system expansion.  The discrepancy between open system expnsion and accretion is an important distinction between the thermodynamics of crystals and liquids.   In section \ref{sec:micro} we make contact with the molecular simulation\cite{swope92,frenkel01,frenkel07,dijkstram12,kofke18}, hydrodynamics\cite{martinpc72,fleming76,fuchs10} and classical Density functional theory\cite{oettel10,fuchs15,fuchs21} of solids. We also comment  on the similarity to the nanothermodynamics of Hill\cite{hill94,hill01} and related statistical mechanics approaches to systems with long range interactions \cite{ruffo09, ruffo14} which have been a guide for the development of the extended thermodynamics in this paper. We conclude in section \ref{sec:concl} with a summary and a list of what is missing. 

\section{Generalized Thermodynamics} \label{sec:thermech}

\subsection{Grand free enthalpy} \label{sec:ggibbs}
Stretching a lattice increases volume $V$  but for an elastic expansion the number of sites remains the same and therefore, in a closed system, also the occupation (particle number $N$ is constant).   Occupation can only be changed by inserting or removing particles keeping $M$ fixed whatever the value of volume.  Site occupation $N/M$ and particle density $N/V$ are distinct intensive variables.  To tell them apart we must treat $M$ as an additional independent extensive variable. This is the basic premise of LC theory. The Gibbs equation for the internal energy is generalized accordingly
\begin{equation}
dU_e =TdS  -pdV + \mu dN + \nu dM
\label{dUext}
\end{equation}
 where $T$ is the temperature and $S$ the entropy.  $p$ is the hydrostatic pressure.  Shear deformation is excluded in this presentation.  The focus is on the effect of volume elasticity.  $\mu$ is the chemical potential of the particles. We allow for empty sites (vacancies) but not for interstitials. Therefore $N \le M$ with $N=M$ corresponding to the ideal solid.  $\nu$ is the new thermodynamic force conjugate to $M$.  The physical interpretation of $\nu$ will become clear in more detailed analysis of the thermodynamics of the LC lattice model.  
 
If not for the last term Eq.~\ref{dUext} would be no different from the familiar Gibbs equation for one-component homogeneous liquids.  This term is a minimal extension required to account for volume strain and vacancies in solids.   These two properties are the fundamental intensive thermodynamic variables distinguishing a solid under hydrostatic pressure from a liquid.  We are therefore led to introduce a new thermodynamic free energy specific for the solid state
\begin{equation}
 \mathcal{E}  = U_e  -TS  + pV - \mu N  
\label{gGibbs}
\end{equation}
Substituting Eq.~\ref{dUext} we have for its differential
\begin{equation}
 d\mathcal{E}  = - SdT + Vdp - N d\mu + \nu dM
\label{dgGibbs}
\end{equation}
$\mathcal{E}$ plays the role of a grand Gibbs free energy quantifying a possible mismatch between the free enthalpy $G_e$ and $\mu N$  or equivalently a great grand potential making up the difference between grand potential $\Omega_e$ and $- pV$. 
\begin{equation}
\mathcal{E} = G_e - \mu N = \Omega_e + pV
\label{mismatch}
\end{equation}
$\mathcal{E} $, also called the null potential, vanishes for a regular liquid with short range interatomic interactions. $\mathcal{E} $ is finite for the self gravitating systems of astronomy and for plasmas\cite{ruffo14}. The microscopic interactions in a hard-sphere or Lennard-Jones solid are the same short range pair forces as in the liquid.  However the correlations enforced by solid rigidity (the lattice) are long range requiring a similar generalization of the macroscopic thermodynamics.  $M$ and $\nu$ must be regarded as configurational quantities (order parameters) characteristic of crystals.  

Constraining particles to lattice sites has further profound thermodynamic consequences. The system size scaling appropriate for liquids fails. Joint increase of volume and  particle number keeping density $N/V$  constant does not lead to a larger system with identical intensive  properties.   The number of lattice sites must be increased by the same ratio. Otherwise the increase in $N$ would lead to a decrease in the density of vacancies.  With  $M$  included in system size scaling extensivity is restored.  The Euler-Gibbs principle  applies and Eq.~\ref{dUext} can be integrated to
\begin{equation}
U_e = TS -pV + \mu N + \nu M
\label{Uext}
\end{equation}
Substituting the definition  Eq.~\ref{gGibbs} we find
\begin{equation}
\mathcal{E} = \nu M
\label{intgdh}
\end{equation}
Eqs.~\ref{Uext} and \ref{intgdh} are general thermodynamic "laws"  for a LC crystal and should not be confused with  constitutive relations which are more specific. An admissible constitutive relation should of course satisfy Eq.~\ref{Uext}. 

Substituting Eq.~\ref{intgdh} in Eq.~\ref{mismatch} the gap between molar Gibbs free energy and the chemical potential can be expressed in terms of $\nu$
\begin{equation}
\dfrac{G_e}{N} - \mu =\frac{ \nu}{ c} 
\label{Gmumis}
\end{equation}
where $c$ is the occupation
\begin{equation}
  c = \frac{N}{M}, \qquad 0 \le c \le 1 
  \label{defc}
  \end{equation}
The maximum value of $c$ is imposed by the single site occupation constraint adopted in LC theory.
The density of the grand potential and the pressure suffers from a similar discrepancy
\begin{equation}
\frac{\Omega_e}{V}   + p  = \frac{\nu}{v^c}
\label{Omapmis}
\end{equation}
$v^c$ is the volume of a lattice cell of the deformed solid. 
\begin{equation}
  v^c = \frac{V}{M}
  \label{defvc}
  \end{equation}
 As will be justified below the thermodynamic force represented by $\nu$ is in general not zero.  $\nu$ differentiates in a most direct way between the solid and liquid phase of a material.  For the example model system outlined in section \ref{sec:model} $\nu$ increases,  rather alarmingly,  approaching the limit of full occupation of the ideal solid.  

\subsection{Gibbs-Duhem relation and absorption} \label{sec:gdhtherm}
The Gibbs-Duhem relation is a corner stone of the thermodynamics of homogeneous liquids.  In view of  Eq.~\ref{Uext} one would expect the solid to add a $M d \nu$ term.  Such a relation can be obtained by substituting the differential of Eq.~\ref{intgdh} in Eq.~\ref{dgGibbs}. The result for an isothermal process ($dT =0$) is
\begin{equation}
M d\nu = V dp - N d \mu
\label{gdhext}
\end{equation}
Following the Gibbs treatment of interfaces,  we next convert Eq.~\ref{gdhext} to the equivalent of an adsorption isotherm.  The partial derivative obtained by setting $dp = 0$ and dividing by $d \mu$  generates the occupation under isobaric isothermal equilibrium. 
\begin{equation}
\left(\frac{ \partial \nu}{\partial \mu}\right)_{p,T}  =  - c
\label{dnudmu}
\end{equation}
The other way around gives the deformed cell volume of Eq.~\ref{defvc} of an open system
\begin{equation}
\left(\frac{ \partial \nu}{\partial p}\right)_{\mu,T}  =  v^c
\label{dnudp}
\end{equation}
$M$ plays the role of the system size determining state variable and is fixed.   

 Eqs.~\ref{dnudmu} and \ref{dnudp} specify two gradients of $\nu$ in the isothermal $(\mu,p)$ plane  equal to  $ -c$ and $v^c$ respectively.  Occupation and cell volume are well behaved positive definite quantities which are accessible to experiment.  $v^c$ can be estimated from the spacing of the maxima in diffraction patterns.  The value of $M$ then follows by dividing the volume of the crystal by $v^c$ as specified by Eq.~\ref{defvc}.   Given $M$,  occupation $c = N/M$ is also fixed.   Relative values of $\nu$ can therefore in principle be obtained by thermodynamic integration.  We can conclude therefore that $\nu$ must assume non-zero values almost everywhere in the $(\mu,p)$ plane,  except perhaps at special nodal curves. 

The argument in this section is greatly inspired by the parallel to Hill's thermodynamics for nanosystems\cite{hill94,bedeaux23}.  Due to the small size the differentials of pressure, temperature and chemical potential of nanosystems are independent violating the standard Gibbs-Duhem equation.  To recover this very useful relation,  Hill constructs an ensemble of replicas of the nanosystem.   This introduces an extensive thermodynamic potential $\mathcal{E}$ which can be compared  to Eq.~\ref{gGibbs} with $M$ acting as the number of replicas.  While the physics is different the thermodynamic equations are formally similar.  In fact Eqs.~\ref{dnudmu} and \ref{dnudp} are virtually copies of the Gibbs-Duhem expressions derived in Ref.~\citenum{hill01}.  Hill's nanothermodynamics has also been the template for the thermodynamics of macroscopic systems with long range interactions developed by Ruffo and coworkers\cite{ruffo09,ruffo14}.  Here the physical connection to LC lattices  is more direct as will be explained in somewhat more detail in section \ref{sec:longrange}.     

We should  add of course that generalized thermodynamic $\left(\mu, p, T\right)$ treatments of fully open system are not new. The theory goes back to Guggenheim\cite{guggenheim85} who, however, did not seem to have definite applications in mind (see also Ref.~\citenum{sack59}).  More recently this was taken up as a challenge for the development of computer simulation methodology\cite{ray93,vrabec23}.

\section{Rigid lattice} \label{sec:rigid}
\subsection{Free energy and chemical potential} \label{sec:mu0}

In the remainder of the paper the thermodynamic theory of section \ref{sec:thermech} will be applied to a simple uniform one-component LC model system under hydrostatic pressure. The critical element of the model is the coupling between occupation and volume stress. The particular formulation of this coupling is not standard.  It is designed to highlight the difference between the hydrostatic pressure in a solid and liquid.  We start therefore with a rigid lattice where there is no such coupling. The system consists again of $M$ sites occupied by $N$ particles.  The Gibbs equation  is 
\begin{equation}
dU_r =TdS + \mu dN + \nu dM
\label{dUr}
\end{equation}
Compared to Eq.~\ref{dUext} the $-pdV$ term is missing but the $\nu M$ term is there.  Similarly Eq.~\ref{Uext} is shortened to
\begin{equation}
U_r = TS + \mu N + \nu M
\label{Ur}
\end{equation} 
The role of thermodynamic volume of the lattice gas is taken over by $M$. The rigid lattice is therefore a good preparation for the molecular understanding of $\nu$.

The free energy $F_r$ is a function of $N$ and $M$ and is a sum of an entropy term $F_s$ and binding energy $F_b$.
\begin{equation}
F_r\left(N,M\right) = F_s\left(N,M\right) + F_b\left(N,M\right)
\label{F0}
\end{equation}
The dependence on temperature is suppressed. In an homogeneous system $F_s$ can be written as a product of the number of sites and an entropy per site
\begin{equation}
F_s \left(N,M\right) = M f_s\left(c\right)
\label{Fs}
\end{equation}
$f_s$ is a function of the occupation $c$ as defined in Eq.~\ref{defc}.  $f_s$ the  Langmuir function familiar from the molecular thermodynamics of adsorption. 
\begin{equation}
 f_s\left(c\right) = k_{\mathrm{B}} T
 \left( c \ln c + \left(1-c \right) \ln \left( 1-c \right)   \right)
 \label{fsc}
 \end{equation}
 $F_b$ is the binding energy of the  particles
\begin{equation}
F_b \left(N, M \right) = - N I\left(c\right)
\label{Fb}
\end{equation}
with $I(c) >0$ the site binding energy.   The dependence on occupation is meant to model the interaction between particles at the mean field level.  Note that the entropy of Eq.~\ref{Fs} is a sum over sites and the binding energy 
of Eq.~\ref{Fb} a sum over atoms. 

The chemical potential $\mu$ is the derivative with respect to the number of particles. The number of sites is fixed.
\begin{equation}
\mu = \left( \frac{\partial F_r}{\partial N} \right)_M = M \left( \frac{\partial f_s}{\partial N} \right)_M - I - N \left( \frac{\partial I}{\partial N} \right)_M
\label{mudef0}
\end{equation}
Applying the chain rule the entropy term is factorized as
\begin{equation}
M \left( \frac{\partial f_s}{\partial N} \right)_M = \frac{d f_s}{d c}
\label{dnchain}
\end{equation}
Substituting Eq.~\ref{fsc} we have 
\begin{equation}
\frac{d f_s}{d c} = k_{\mathrm{B}} T \ln \left( \frac{c}{1-c} \right)
\label{dfsc}
\end{equation}
Similarly for the binding energy derivative in  Eq.~\ref{mudef0} 
\begin{equation}
N \left( \frac{\partial I}{\partial N} \right)_M = c \frac{dI}{d c} 
\label{dIc}
\end{equation}
Defining is the response function
\begin{equation}
\gamma_r(c) = - c  \frac{dI}{d c} = - c I^\prime(c)
\label{defgamr}
\end{equation}
we can write for the chemical potential
\begin{equation}
\mu = k_{\mathrm{B}} T \ln \left( \frac{c} {1-c}\right) - I  + \gamma_r
\label{mu0}
\end{equation}
$\gamma_r > 0$ for repulsive interactions and $\gamma_r <0$ for attractive interactions.

Setting $\gamma_r = 0$ we recover the well-known non-interacting  Langmuir adsorption model 
\begin{equation}
c_0  = \frac{ \lambda}{ \lambda +1 }
\label{cmu0}
\end{equation}
with the activity $\lambda$ given by
\begin{equation}
\lambda = \exp  \left[  \frac{\left(\mu +I \right)}{k_{\mathrm{B}} T} \right]
\label{lamb0}
\end{equation}
The density of unoccupied sites (vacancies)
\begin{equation}
h = 1 -c = \frac{M-N}{M}
\label{defh}
\end{equation}
is in the noninteracting limit
\begin{equation}
h_0 = \frac{ 1}{ \lambda  +1 }
\label{hmu0}
\end{equation}
Finally we remind the reader of the singular behaviour of Langmuir entropy approaching full occupation ($ c \rightarrow 1$).  The entropy per site $f_s(c)$ of Eq.~\ref{fsc} vanishes in this limit ($f_s(1) = 0$).  Its derivative Eq.~\ref{dfsc} (the configurational chemical potential) diverges to $+ \infty$.  The opposite limiting case of an empty lattice ($c \rightarrow 0$) shows a similar singularity.

\subsection{Lattice site potential} \label{sec:nu0}

The thermodynamic force conjugate to the lattice site number $M$ is obtained by evaluating the derivative of free energy with respect to $M$ for a fixed number of particles. 
\begin{equation}
\nu = \left( \frac{\partial F_r}{\partial M} \right)_N  = f_s + M \left( \frac{\partial f_s}{\partial M} \right)_N - N \left( \frac{\partial I}{\partial M} \right)_N
\label{nudef}
\end{equation}
As in Eq.~\ref{dnchain} we use the chain rule
\begin{equation}
M \left( \frac{\partial f_s}{\partial M} \right)_N =- c \frac{d f_s}{d c} 
\end{equation}
Similarly for the site binding term in Eq.~\ref{nudef}. Combining using Eq.~\ref{dfsc} yields
\begin{equation}
\nu  = f_s -  k_{\mathrm{B}} Tc  \ln \left( \frac{c}{1-c} \right) - c \gamma_r
\label{nu0}
\end{equation}
with $\gamma_r$ defined in Eq.~\ref{defgamr}.
Substituting the expression for $f_s$  given in Eq.~\ref{fsc} and working out the sum we find that a number of terms cancel leaving us with
\begin{equation}
\nu = k_{\mathrm{B}} T  \ln \left( 1 - c \right) - c \gamma_r
\label{nuh0}
\end{equation}
which is complementary to Eq.~\ref{mu0} for $\mu$.

What to make of this quantity which we haven't even given a proper name yet?  It is tempting to think of $\nu$ as the chemical potential of the lattice sites.  However,  let us investigate a sparsely occupied non-interacting system ($N \ll M, \; \gamma_r = 0$).  Expanding in $c$ and multiplying by $M$ gives in lowest order
\begin{equation}
 \nu M  = - N k_{\mathrm{B}} T 
 \label{pideal}
\end{equation} 
Eq.~\ref{pideal} is the ideal gas law for a lattice gas and is easily verified by Monte Carlo simulation. This suggests that $\nu$ is more of a stress (negative pressure) than a chemical potential.  This interpretation is supported by the Gibbs-Duhem relation for the rigid lattice derived from the combination of  Eqs.~\ref{dUr} and \ref{Ur}
  \begin{equation}
   d \nu = - c d \mu
   \label{gdh0}
 \end{equation}
 Eq.~\ref{gdh0} is the lattice gas equivalent of the celebrated $dp = \rho d \mu$ equation for a ``off-lattice''  liquid of density $\rho$.   
 
 The regime of interest to us is  the nearly complete lattice. Here  Eq.~\ref{nuh0} behaves rather differently.  The concentration of vacancies $h = 1- c$ is small and the logarithmic term dominates.  Approximating the interaction term $- c \gamma_r$ by its limiting value  $ I^\prime (1) $ at  $c=1$  (see Eq.~\ref{defgamr}) we have
 \begin{equation}
\nu =  k_{\mathrm{B}} T  \ln h +I^\prime(1)
\label{nuh0c1}
\end{equation}
 Now, instead of a pressure,  a chemical potential picture seems more fitting. We opt therefore calling $\nu$ by the more neutral name of "lattice site potential".  
 
 Finally, we compare to expression Eq.~\ref{mu0} for the chemical potential evaluated in the same $c \rightarrow 1$ approximation.
 \begin{equation}
 \mu = - k_{\mathrm{B}} T \ln h - I(1)  - I^\prime(1)
 \label{muh0c1}
 \end{equation}
 The $ k_{\mathrm{B}} T  \ln h$ term appears also here. Indeed there are two distinct ways of creating vacancies. The chemical way is removing a particle from a lattice with a given number of sites (constant $M$).  The change in free energy per particle is $-\mu$ as given by Eq.~\ref{muh0c1}.  The loss of binding energy is offset by the increase in entropy (recall $I > 0$ in the convention of Eq.~\ref{Fb}).  Alternatively we can add a lattice site keeping the number of particles fixed  (constant $N$). The change in free energy is given by Eq.~\ref{nuh0}. There is a similar gain in entropy without the direct penalty of binding energy. The number is particles is conserved.  Still there is a relatively minor change in interaction energy due to the decrease in density. This is also the origin of the last term in Eq.~\ref{muh0c1}. Note that new sites for a regular lattice gas/Ising system are not inserted but  added at the periphery. This detail will become crucial later for the compressible lattice.

\subsection{Absorption isotherm and accretion} \label{sec:accr0}

 Returning to   Eq.~\ref{gdh0} formulated in terms of derivatives is this equation is the Gibbs absorption isotherm for the Langmuir model.
  \begin{equation}
\left( \frac{ \partial \nu}{ \partial \mu}\right)_T  = - c
\label{dnudmuc}
 \end{equation}
 This equation not only applies to a noninteracting system  but is  a generally valid thermodynamic identity for the lattice gas.  It can be checked easily for $\gamma_r = 0$ which we will do below.   This exercise gives  further insight introducing an important quantity needed for later reference,  namely  the  isothermal susceptibility
 \begin{equation}
\chi = \left(\frac{ \partial c}{ \partial \mu}\right)_T
\label{defchi}
\end{equation}
Substituting expression Eq.~\ref{cmu0} we see that $\chi$ for the noninteracting system is a product of vacancy and particle population
\begin{equation}
\chi_0 =  \frac{ \beta \lambda}{ \left(\lambda +1 \right)^2} =  \beta h_0c_0
\label{chihc}
\end{equation}
where in the usual notation $\beta = 1/k_{\mathrm{B}}T$.  The susceptibility vanishes in the limit of a  full as well as empty lattice as is characteristic for  single site occupancy. Applying the chain rule (suppressing the constant $T$ condition)
\begin{equation}
\frac{ \partial \nu}{ \partial \mu} = \chi \frac{ \partial \nu }{ \partial c} 
\label{dnudmuchi}
\end{equation}
Then evaluating the occupation derivative using the expression for $\nu$ of Eq.~\ref{nuh0} leaving out the $\gamma_r$ term we obtain
\begin{equation}
\frac{\partial \nu}{\partial c} = - \frac{k_{\mathrm{B}}T}{ h}
\end{equation}
Substituting in Eq.~\ref{dnudmuchi} with  Eq.~\ref{chihc} recovers the Gibbs-Langmuir adsorption relation 
 Eq.~\ref{dnudmuc}.

Adding lattice sites at the surface and occupying them with additional particles leaving the same fraction of empty sites as in the bulk crystal is a physical process.  This amounts simply to enlarging the crystal by accretion. The infinitesimal increase in $M$ and $N$ are related as $dN = cdM$. The corresponding increase in free energy is found from the Gibbs free energy equation
\begin{equation}
dF_r =  \nu dM +  \mu dN  = \left(\nu + c \mu \right) dM
\label{dF0munu}
\end{equation}
Substituting expressions Eq.~\ref{mu0} and  Eq.~\ref{nu0} for  $\mu$ respectively $\nu$ and comparing to Eq.~\ref{fsc} we find
\begin{equation}
\nu + c \mu = f_s - c I
\label{numu0}
\end{equation}
As expected the increase in free energy for accretion is proportional  free energy per lattice site 
\begin{equation}
dF_0 = \left( f_s - c I \right) dM
 \label{dF0dV}
\end{equation}
Multiplying Eq.~\ref{numu0} by $M$ 
\begin{equation}
F_0 = N \mu + M \nu
\label{euler0}
\end{equation}
we recognize the Euler equation for the rigid lattice.  With $M$ included as an additional extensive  thermodynamic variable the free energy is extensive.

\section{Compressible lattice model} \label{sec:compress} \label{sec:model}

  The bonds connecting lattice sites in a compressible LC model are flexible.   However,  the basic structure of a lattice gas is retained.  Lattice sites are separate structural elements which continue to exist whether they are occupied or empty. Therefore,  also the bonds to a site are not broken if a particle occupying the site leaves.   At this level of approximation the mechanical response  is modeled by an elastic energy depending on strain treated as an independent state variable.  In our simple model of a uniform system under hydrostatic pressure deformation is restricted to isotropic expansion or contraction of volume $V$.  The number of lattice sites is conserved\cite{cahn73}.  This is the fundamental characteristic of reversible elastic response.  $M$ remains therefore an independent extensive variable in addition to particle number $N$ as it was for the rigid lattice.   The new variable compared to the rigid lattice is therefore volume $V$.
 
\begin{figure}
\includegraphics[width=0.9\columnwidth, clip=true]{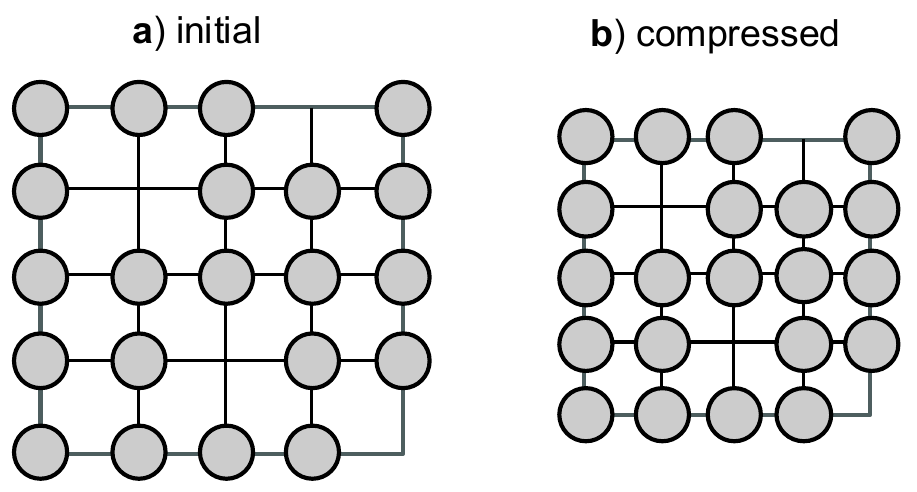}
\caption{\label{fig:Ibdens} Lattice gas with sites occupied by particles of finite atomic radius.   The initial state {\bf a} is the  rigid lattice defined in  section \ref{sec:rigid}. Particle interactions in the mean field approximation are functions of occupation $c = N/M$ (Eq.~\ref{Fb}).  Compression  reduces particle separation distance without changing the atomic radius as shown in {\bf b}.  At the mean field level this is accounted for by making the site binding energy depend on the particle density $\rho = N/V$ in the deformed lattice (Eq.~\ref{Fbrho}). }
\end{figure}

 The Helmholtz free energy at a given fixed temperature is determined by the values of $N, V$ and $M$ and resolved in the sum of three contributions
\begin{equation}
F_e\left(N,V,M\right) = F_s\left(N,M\right) + F_b\left(N,V\right) + F_j\left(V,M\right)
\label{Fe}
\end{equation}
 $F_s$ is the Langmuir entropy defined in Eqs.~\ref{Fs} and Eq.~\ref{fsc}.  The second term in Eq.~\ref{Fe} is a modification  of the binding energy $F_b$ of Eq.~\ref{Fb}.  The function $I$ is the same,  the argument however is now not occupation but deformed density
 \begin{equation}
F_b\left(N,V\right) = - N I\left(\rho\right)
\label{Fbrho}
\end{equation}
with $\rho$ defined as usual by
\begin{equation}
 \rho =  \frac{N}{V} 
 \label{defrho}
\end{equation}
the justification for replacing $c$ by $\rho$   is that interaction parameters such as the size of the particles are invariant under deformation,  only the distance between particles is affected. (see Fig.~\ref{fig:Ibdens}).  

The third term in Eq.~\ref{Fe} is the deformation energy which we will write as
\begin{equation}
F_j\left(V,M \right)  = M  g_j\left(J\right)
\label{Fjdef}
\end{equation}
with $g_j$ representing the elastic energy per site.   $g_j$ is a function of the stretch $J$ defined as
\begin{equation}
J = \frac{V}{V_R} 
\label{JV}
\end{equation}
where $V_R$ is the volume of a stress free reference state.  The connection to the thermodynamics of section \ref{sec:thermech} is that $V_R$ is an extensive quantity scaling with  $M$
\begin{equation}
V_R = v_R^c M
\label{VRM}
\end{equation}
 The prefactor $v^c_R$ is the volume of a lattice cell in the reference state and is a material constant.  $J$ is therefore an intensive variable proportional to the deformed cell volume
  $v^c$ of Eq.~\ref{defvc} 
 \begin{equation}
J = \frac{V}{M v_R^c} = \frac{v^c}{v_R^c}
\label{Jvcell}
\end{equation}
 We will use the simplest harmonic approximation for $g_j$
\begin{equation}
g_j\left(J\right) = \frac{\alpha_j}{2} \left(J-1\right)^2 = \frac{\alpha_j}{2} \epsilon^2 
\label{gjhar}
\end{equation}
where $\alpha_j > 0$ acts as a spring constant.   $\epsilon$ is the volume strain related to the volume stretch as 
\begin{equation}
J = 1 + \epsilon
\label{vstrain}
\end{equation} 
Conform the usual definition $\epsilon < 0$ for compression and $\epsilon > 0$ for expansion.

Substituting Eqs.~\ref{Fs},  Eq.~\ref{Fbrho} and Eq.~\ref{Fjdef} for the three terms making up the free energy  Eq.~\ref{Fe} we now have
\begin{equation}
F_e \left(N,V,M\right)= M f_s\left(c\right) - N I\left(\rho\right) + M g_j\left(J\right)
\label{Fhelm}
\end{equation}
The  variables $c, \rho$ and $J$ on the right hand site are each ratios of the extensive state variables  $N,V,M$ in the argument of $F_e$ on the left hand side.  Eq.~\ref{Fhelm} for the free energy is therefore in a form  manifestly linear in system size.  According to the thermodynamic theory of section \ref{sec:thermech} we should therefore be able to eliminated one of the three densities expressing it in terms of the remaining two.  Our selection  for the dependent variable is the number density writing it as a function of $c$ and $J$  using Eqs.~\ref{defc} and \ref{Jvcell}. 
 \begin{equation}
\rho = \left(\frac{N}{M}\right)\left(\frac{M}{V}\right) = \frac{c}{J v_R^c}
 \label{rho2c}
\end{equation}
Factoring out $M$   in Eq.~\ref{Fhelm}
\begin{equation}
F_e \left(N,V,M\right) = M f_e\left(c, J\right)
\label{Fesite}
\end{equation}
we obtain a free energy per site $f_e$  as a function of occupation $c$ (Eq.~\ref{defc}) and stretch $J$ (Eq.~\ref{JV}). 
\begin{equation}
 f_e\left(c,J\right) = f_s\left(c\right) + f_b\left(c,J\right) + g_j\left(J\right)
 \label{fesum}
\end{equation}
 $f_s\left(c\right)$ is the Langmuir entropy per site given in Eq.~\ref{fsc}. $g_j\left(J\right)$ is the elastic energy per site defined in 
 Eq.~\ref{gjhar}.  $f_b\left(c,J\right)$ is the chemomechanical coupling term derived from the density dependent binding energy Eq.~\ref{Fbrho}
 \begin{equation}
  f_b\left(c,J\right) =- c I\left(\rho\right) =- c  I \left( \frac{c}{v_R^c J} \right)
  \label{fbdef}
 \end{equation}
where in the second identity  the deformed density $\rho$  was converted to a function of $c$ and $J$ using Eq.~\ref{rho2c}.  

\section{Thermodynamic derivatives} \label{sec:dtherm}

\subsection{Chemical potential} \label{sec:mue}
 Having decided on a model for the free energy  (Eq.~\ref{Fhelm}) we next determine the chemical potential $\mu$,  pressure $p$ and lattice site potential $\nu$.   Applying the thermodynamics of section \ref{sec:thermech} 
 the differential of $F_e$ is 
 \begin{equation}
dF_e = - pdV + \mu dN + \nu dM
\label{dFext}
\end{equation}
Accordingly,  expressions for the thermodynamic forces are obtained as partial derivatives with respect to the extensive variables $N,V$ and $M$.   The common practice  in material science literature is different (see eg XX).  
The expression for chemical potential and pressure are obtained by differentiating Eq.~\ref{Fesite} to densities $c$ respectively  $J$.    Lattice potentials,  although defined,  are usually not explicitly taken into account (see for example the comprehensive Voorhees Johnson review Ref.~\citenum{voorhees04}).  If needed, a convenient way for the determination of $\nu$ in this intensive variable based scheme is using  Eq.~\ref{gGibbs} for the grand Gibbs free energy per site.
 
Starting with $\mu$ we differentiate $F_e$  with respect to $N$ at fixed volume and number of lattice sites
\begin{equation}
\mu = \left( \frac{ \partial F_e}{\partial N} \right)_{V,M} =  \left( \frac{ \partial F_s}{\partial N} \right)_{M} +   \left( \frac{ \partial F_b}{\partial N} \right)_{V}
\label{mudefe}
\end{equation}
There is no contribution from the $F_j$ term because the elastic energy Eq.~\ref{Fjdef}  is independent of $N$. We have also simplified the notation taking into account that the entropy term is invariant for changes in $V$ and similarly  the binding energy for changes in $M$.  The $F_s$ derivative is the same as in section \ref{sec:mu0}.
\begin{equation}
\left( \frac{ \partial F_s}{\partial N} \right)_{M} = M \left( \frac{\partial f_s(c)}{\partial N} \right)_M
= k_{\mathrm{B}} T \ln \left( \frac{c}{1-c} \right) 
\end{equation}
The derivative of the binding energy adds two terms 
\begin{equation}
 \left( \frac{ \partial F_b}{\partial N} \right)_{V} =
 - I - N \frac{d I}{d \rho} \left(\frac{\partial \rho}{\partial N}\right)_V =
 -I + \gamma
\end{equation}
 where $\gamma = \gamma\left(\rho\right)$ is the binding energy response function defined as
 \begin{equation}
 \gamma\left(\rho\right) = -  \rho \frac{d I\left(\rho\right)}{d \rho} 
 \label{gammadef}
 \end{equation}
 Sign conventions are such that $\gamma > 0$ for repulsive interactions.  Combining we obtain for the chemical potential
\begin{equation}
\mu =  k_{\mathrm{B}} T \ln \left( \frac{c}{1-c} \right) - I + \gamma
\label{mue}
\end{equation}

\subsection{Lattice site potential} \label{sec:nue}
Next is the modification of the lattice potential of Eq.~\ref{nu0} which now consists of two terms
\begin{equation}
\nu = \left( \frac{ \partial F_e}{\partial M} \right)_{N,V} =  \left( \frac{ \partial F_s}{\partial M} \right)_{N} +  \left( \frac{ \partial F_j}{\partial M} \right)_{V}
\label{nudefe}
\end{equation}
$F_s$ is a reference frame free energy  insensitive to changes in $V$. For this contribution Eq.~\ref{nu0} still applies
\begin{equation}
\left( \frac{ \partial F_s}{\partial M} \right)_{N} = k_{\mathrm{B}} T \ln \left( 1 - c\right) 
\end{equation}
The elastic term in Eq.~\ref{nudefe} is new
\begin{equation}
\left( \frac{ \partial F_j}{\partial M} \right)_{V}   =
  g_j + M \frac{d g_j}{d J} \left(\frac{ \partial J}{\partial M} \right)_V 
 \end{equation}
  $d g_j /d J$ is the volume stress which will be indicated by $\sigma_j$.  In the harmonic approximation Eq.~\ref{gjhar} the volume stress is linear in the volume strain
  \begin{equation}
  \sigma_j = \alpha_j \left( J -1 \right) = \alpha_j \epsilon
  \label{sigmaj}
  \end{equation}
  Then with Eq.~\ref{JV}
\begin{equation}
 M \left(\frac{ \partial J}{\partial M}\right)_V = -J
 \end{equation}
and we obtain
\begin{equation}
\left( \frac{ \partial F_j}{\partial M} \right)_{V} =   g_j - J \sigma_j
\end{equation}
In the harmonic approximation these two terms can be added by analytic summation
\begin{equation}
g_j - J \sigma_j = - \frac{\alpha_j}{2} \left(J^2 -1 \right) = - \tfrac{1}{2} \left(J+1 \right) \sigma_j 
\label{virialj}
\end{equation}
which is positive under compression $J<1$.  Gathering terms we obtain
\begin{equation}
\nu =  k_{\mathrm{B}} T \ln \left( 1 - c \right) -  \tfrac{1}{2} \left(J+1 \right) \sigma_j 
\label{nue}
\end{equation}
Note that $\nu$ is not affected by changes in the site binding energy. $\mu$  of Eq.~\ref{mue} shows the complementary behavior of being insensitive to elastic energy. These two energies are combined in the expression for the pressure as we see in the next section.

The default approach  to continuum mechanics of crystals is linear elasticity.  Linearization is straightforward for the lattice site potential of Eq.~\ref{nue}. The second term is to first order in the strain  reduced to the elastic stress $\sigma_j$ of Eq.~\ref{sigmaj}. 
 Changing the $1-c$ argument of the logarithm to the vacancy population $h$ we can write Eq.~\ref{nue} as
\begin{equation}
\nu =  k_{\mathrm{B}} T \ln h - \alpha_j \epsilon
\label{nusigma}
\end{equation}
Comparing to the expression of Eq.~\ref{nuh0} of the rigid lattice we see that the binding energy parameter $\gamma$ has been replaced by the volume stress.  Molecular pressure as defined in Eq.~\ref{pcstar}  has no direct effect on $\nu$.

\begin{figure}
\includegraphics[width=0.9\columnwidth, clip=true]{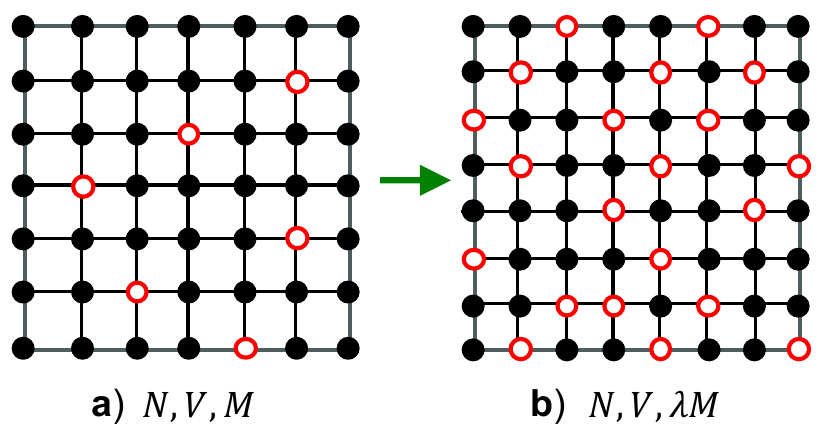}
\caption{\label{fig:minsert} Increasing the number of lattice sites $M$ keeping particle number $N$ and volume $V$ fixed.  {\bf a} shows the initial state and {\bf b} the lattice with $M$ a factor $\lambda$ larger.  This operation changes occupation $c= N/M$ and cell volume $v^c = V/M$ in {\bf a} to  $c/\lambda$ and $v^c/\lambda$ in {\bf b} establishing a balance between vacancy population $h = 1 -c$ and volume stress $\alpha_j \left(J-1\right)$ controlled by lattice potential $\nu$ (Eq.~\ref{nusigma}).}
\end{figure}

Eq.~\ref{nusigma} establishes a correlation  between vacancy concentration and elastic rigidity.  Both quantities are unique to crystals and so is the mechanism for the coupling between the two as illustrated in Fig.~\ref{fig:minsert}.   Inserting lattice sites at fixed  particle number inevitably must create vacancies. (Fig.~\ref{fig:minsert}).  However the definition of $\nu$ of Eq.~\ref{nudefe}  requires that also volume is kept constant in the process forcing cell volume to shrink at the expense of generating volume stress.  The competition between chemical potential of vacancies and volume stress is controlled by the value of $\nu$.   Note that a zero value of $\nu$ can only be realized under compression.  The reason is that the effective chemical potential for vacancies (the $\ln h$) ) cannot be positive and a compensating strain must therefore also be negative. Tension is only possible for finite $\nu$.  This suggest  that $\nu$ in general is not zero as is also indicated by the Gibbs-Duhem relations Eqs.~\ref{dnudmu} and \ref{dnudp}.  The $\mu$ and $p$ derivatives are equal to $-c$ and $v_c$ which  are well behaved densities in stable crystals.  This issues is however controversial as will be discussed in section \ref{sec:molsim}.

\subsection{Elastic and molecular pressure} \label{sec:pressure}
The $V$ derivative of free energy is the hydrostatic  pressure and consists again of two distinct contributions
\begin{equation}
p = - \left( \frac{ \partial F_e}{\partial V} \right)_{N,M} =  -\left( \frac{ \partial F_b}{\partial V} \right)_{N} -  \left( \frac{ \partial F_j}{\partial V} \right)_{M}
\label{pdef}
\end{equation}
The first term is the pressure generated by the density dependence of the binding energy
 \begin{equation}
 \left( \frac{ \partial F_b}{\partial V} \right)_{N} =-  N \frac{d I}{d \rho} \left(\frac{\partial \rho}{\partial V}\right)_N
   = \rho^2 \frac{d I}{d \rho}
  =  - \rho \gamma
\end{equation}
with $\gamma$ the binding energy response function Eq.~\ref{gammadef}.

The second term accounts for the lattice elasticity
\begin{equation}
\left( \frac{ \partial F_j}{\partial V} \right)_{M}   = M  \frac{d g_j}{d J} \left(\frac{ \partial J}{\partial V} \right)_{V_R} 
=  \frac{ \alpha_j}{ v^c_R } \left(J -1 \right)
\end{equation}

Changing over to the scaled pressure  defined as
\begin{equation}
p^\star = v_c^R p
\label{defpstar}
\end{equation} 
we can write the pressure as the difference 
\begin{equation}
p^\star = p_c^\star - \sigma_j
\label{pstar}
\end{equation}
of a molecular pressure
\begin{equation}
p_c^\star \left( \rho \right) = \gamma \left(\rho\right) \left(\frac{\rho}{\rho_R} \right)
\label{pcstar}
\end{equation}
 and  the elastic stress $\sigma_j$ of Eq.~\ref{sigmaj}.  $\rho_R = 1/v_R^c$ is the reference density.  $p_c^\star$ and $\sigma_j$ are physically distinct contributions to the pressure.  For an undeformed system $\sigma_j = 0$.  The elastic volume stress vanishes.   However the molecular pressure remains finite..    The the general constitutive framework of  Eq.~\ref{Fe}  allows by construction for an unstrained state under hydrostatic pressure. (see further section X).

\section{Quadratic occupation approximation} \label{sec:epstrain}

\subsection{Minimal model for the site binding energy} \label{sec:cstrainint}
 To complete the specification of the free energy we would have to choose a model for the function $f_b$ of Eq.~\ref{fbdef}.  We will not go that far in this investigation. The target is stable crystal with a characteristically low concentration of vacancies which can be treated by perturbative approach.   The binding energy function $I(\rho)$ is linearized using the the fully occupied crystal under zero elastic stress as a reference.  The density in this state is simply $\rho_R = 1/v_R^c$.  Expanding $I\left(\rho\right)$ to first order we have
\begin{equation}
I\left(\rho\right) =I_b +\gamma_b - \frac{ \gamma_b  \rho}{\rho_R} 
\label{Irhon1}
\end{equation}
$I_b = I\left(\rho_R \right)$ is the energy for removing a particle from a vacancy free undeformed crystal.  
The coefficient $\gamma_b = \gamma\left(\rho_R\right) $ is the corresponding value of the density response function  of Eq.~\ref{gammadef}.
Substituting in Eq.~\ref{fbdef}  gives for the occupation strain coupling
\begin{equation}
   f_b\left(c,J\right)  = - c \left(I_b + \gamma_b \right) + \frac{\gamma_b c^2}{J} 
\label{fbj}
\end{equation}
$I_b$ and $\gamma_b$ are constants.  The occupation strain coupling is a quadratic function of  occupation. 

The expression for the chemical potential in the quadratic occupation approximation is readily obtained by substituting Eq.~\ref{Irhon1} in Eq.~\ref{mue} noting that the corresponding linear approximation for 
$\gamma\left(\rho\right)$ is 
\begin{equation}
\gamma\left(\rho\right) = \frac{\gamma_b \rho}{\rho_R}  
\label{gamma1}
\end{equation}
The result is
\begin{equation}
\mu = - k_{\mathrm{B}} T \ln \left( \frac{1-c}{c} \right) - I_b - \gamma_b   + 2  \gamma_b \left(\frac{ c}{J}\right)
\label{mue1}
\end{equation}
The molecular pressure Eq.~\ref{pcstar} is quadratic to lowest order in  $c/J$ 
\begin{equation}
p^\star_c = \gamma_b \left(\frac{c}{J}\right)^2  
\label{pcstarj}
\end{equation}
Substituting in  Eq.~\ref{pstar}  gives for the full pressure 
\begin{equation}
p^\star =  \gamma_b \left(\frac{c}{J}\right)^2 - \alpha_j \left( J - 1 \right)
\label{pstarj}
\end{equation}
Expression Eq.~\ref{nue} for the lattice potential is carried over without change to the quadratic interaction approximation. 

\subsection{Accretion of the compressible lattice} \label{sec:accre}

As a first application we will revisit the problem of accretion.  Viewed as a formal equilibrium thermodynamic process accretion consists of  enlarging  the system by equally scaling up all extensive state variables.  For an extensive system this amounts to adding more of the same and the free energy should increase in proportion.
This was verified in section \ref{sec:accr0} for the rigid lattice gas model by carrying out an actual accretion operation.  It was meant as an illustration of what this mysterious  $\nu$ stands for and why it cannot be ignored.
  We will go through the same exercise for the compressible lattice now also scaling up volume. 

The joint increase of $M$ and $N$  proceeds as in section \ref{sec:accr0}.  The lattice is extended by $d M$ sites simultaneous with insertion of $dN = c dM$ particles to maintain constant occupation. The corresponding increase in free energy is given by Eq.~\ref{dF0munu}.  Substituting Eq.~\ref{mue1} for $\mu$ and Eq.~\ref{nue} for $\nu$ we obtain for the increase of free energy per lattice site
\begin{equation}
 \nu + c \mu  =  f_s  -c \left( I_b + \gamma_b \right) +2 J  p_c^\star - \tfrac{1}{2} \left(J+1 \right) \sigma_j 
\label{numup}
\end{equation}
Comparing to  Eq.~\ref{numu0} for the rigid lattice there are now  additional terms related to the molecular pressure and volume stress.  The explanation of these mechanical energies is that the increase in $M$ and $N$ is carried out at fixed volume which changes the stretch $J \propto V/M$.  However,  the stretch is an intensive 
and  must be kept constant as well during enlargement and we must scale up volume accordingly. 

The increment in volume consistent with an increase in $M$ keeping $J$ fixed is found by inverting Eq.~\ref{JV}
\begin{equation}
dV = J dV_R = J v_R^c dM
\end{equation}
Multiplying with the pressure gives the corresponding change in mechanical energy
\begin{equation}
p dV = p J v_R^c dM = p^\star J dM  
\end{equation}
where Eq.~\ref{defpstar} was used in the second step. Substituting Eq.~\ref{pstarj} for the pressure and adding to Eq.~\ref{numup} we obtain
after some rearranging
\begin{align}
\nu + c \mu  - p^\star J =  & f_s  -c \left( I_b + \gamma_b \right)  + J  p_c^\star 
\nonumber \\ & -  \tfrac{1}{2} \left(J+1 \right) \sigma_j  + J \sigma_j 
\end{align}
Combining the second and third term on the rhs we recover the binding energy $f_b$ of Eq.~\ref{fbj}
\begin{equation}
- c \left( I_b + \gamma_b \right) + J p_c^\star = f_b
\end{equation}
Similarly  the last two terms add up to the elastic energy $g_j$ applying Eq.~\ref{virialj}. Gathering terms we find
\begin{equation}
\nu + c \mu  - p^\star J   = f_s  + f_b + g_j
\label{fegdh}
\end{equation}
Multiplying by $dM$ using Eq.~\ref{Fesite} delivers the identity we have been looking for
\begin{equation}
\left(\nu + c \mu  - p^\star J\right) d M   = f_e dM
\label{Fexte}
\end{equation}
The (isothermal) work for a consistent uniform change of each of the extensive variables adds up to the expected change in total free energy.   This energy balance could not have been satisfied if the $\nu$ term had been omitted. 


\subsection{Eigenstrain  and reference for deformation} \label{sec:selfstress}

The constitutive model of section \ref{sec:model}  separates pressure  in a density dependent molecular pressure and elastic stress (Eq.~\ref{pstar}).  This raises questions about definition and interpretation of the reference state for the strain.  Normally the reference is a state with zero stress.  However setting $p^\star = 0$ in Eq.~\ref{pstar} imposes a balance between elastic stress $\sigma_j$ and molecular pressure $p_c^\star$ allowing both to be non-zero.  Denoting the density at zero pressure by $\rho_0$ and the stretch by $J_0$ and recalling that molecular pressure is determined by density only we have
\begin{equation}
\alpha_j \left(J_0 -1 \right) = p_c^\star\left(\rho_0\right) 
\label{zeropeq}
\end{equation}
$p_c^\star(\rho_0)$ will in general nonzero leaving us with a finite residual strain $\epsilon_0 = J_0 -1 $.  Following the terminology of the continuum mechanics of defects $\epsilon_0$ will be referred to as eigenstrain\cite{eshelby56,eshelby75,gurtin00,kienzler00}.  .   

It may seem  that  state with zero total pressure is not suitable as a reference for elastic deformation.  However according to the thermodynamics of section \ref{sec:thermech},  in contrast to a liquid,  a crystal under open boundary conditions has a second independent degree of freedom in addition to pressure., namely the chemical potential.   This suggests that it is in principle  possible to vary the chemical potential under constant zero pressure searching to eliminate elastic strain.  However the molecular pressure would also vanish in this state.   It is not at clear that this condition can be realized for a given system.  An example where this fails is a system with repulsive interactions.  A more practical choice is therefore a zero elastic stress state with finite total pressure equal to the molecular pressure.

 Such a finite pressure reference will be worked out now for the model of section \ref{sec:model} in the quadartic occupation approximation of section \ref{sec:cstrainint}.  The expression for the pressure in this approximation is given by Eq.~\ref{pstarj}.  
 Replacing $c/J$ by $\rho/\rho_R$  using Eq.~\ref{rho2c}  gives
\begin{equation}
 p^\star = \gamma_b  \left(\frac{\rho}{\rho_R} \right)^2 - \alpha_j \left( J -1 \right)
 \label{pstarrho}
\end{equation}
which leads to the model specific formulation of Eq.~\ref{zeropeq} 
\begin{equation}
J_0 - 1 = \frac{\gamma_b}{\alpha_j}  \left(\frac{\rho_0}{\rho_R} \right)^2
\label{eps0}
\end{equation}
 As could be anticipated the eigenstrain $\epsilon_0 = J_0 - 1$ is proportional to the occupation-strain coupling parameter $\gamma_b$ and will be larger for softer systems with small $\alpha_j$.   

Pressure in a strain free state for this model is written as  
 \begin{equation}
p^\star = \gamma_b  \left(\frac{\rho}{\rho_R} \right)^2
\label{pstarR}
\end{equation}
Again because of number of state variables of a solid is one more compared to a liquid the density $\rho$ in Eq.~\ref{pstarR} is arbitrary.   We fix $\rho$ by requiring that the reference for deformation coincides with the more strict reference for the Taylor expansion of $I\left(\rho\right)$ in section \ref{sec:cstrainint}.     There we  assumed  that in addition to $J=1$ also $c=1$ (full occupation).   In this limit $\rho = \rho_R$ by definition.  The pressure $p_R$ in the ideal undeformed crystal is in the approximation of Eq.~\ref{pstarR} simply equal to $\gamma_b$.   $p_R$ is positive for repulsive interactions $\gamma_b > 0$ while system is under tension ($p_R < 0$) for attractive interactions ($\gamma_b < 0$).

 The mechanism coupling occupation to strain developed in this work is  different  from the original chemomechanical interaction in LC theory.    LC make no distinction between molecular pressure and elastic stress.  All stress is elastic and determined for small deformation by the standard constitutive relation of linear elasticity theory.   Restricted to isotropic volume dilatation this amounts to $\sigma_j = \alpha_j  \epsilon$
 in our notation.    Instead LC resolve strain in an elastic and composition component\cite{cahn73,cahn85}.    Formulated in terms of the inverse strain stress relation this amounts to
\begin{equation} 
\epsilon = 3 \kappa_j \sigma_j + 3 \eta c
\label{compstrain}
\end{equation} 
 where $\kappa_j = 1/\alpha_j$ is the bare lattice compressibility.   The coefficient  $\eta$ is the LC chemomechanical coupling parameter.  
 
 Similar to Eq.~\ref{pstarrho} not all strain is eliminated under zero external pressure.   The remaining strain obtained by substituting $\sigma_j= 0$ in Eq.~\ref{compstrain} is
 \begin{equation}
 \epsilon_0 = 3 \eta c
 \label{compstrain0}
 \end{equation}
 which again can be interpreted as eigenstrain and compared to Eq.~\ref{eps0}.  This comparison is particularly concise when evaluated for our reference state. 
 \begin{equation}
\eta = \tfrac{1 }{3}\gamma_b \kappa_j 
\label{etacm}
 \end{equation}
 As required by consistency,  the composition strain parameter $\eta$ interpreted in our formalism vanishes for $\gamma_b = 0$.  Moreover for hard system  the effect is only minor.
 
 Evidently,  additive decomposition of the pressure leads to factorization of the composition strain of LC theory. 
 While this may be  an appealing picture, it is questionable whether the parallel between LC composition  strain and molecular pressure as defined here is physically justified.  LC composition strain is meant to model the deformation of alloys due to variation in composition (Vegard's law). Our model system is however a single-component crystal with a small concentration of vacancies.  The eigenstrain of Eq.~\ref{eps0} is still finite even in an ideal crystal without vacancies.

\subsection{Small deformation and pseudo ideal limit} \label{sec:smalleps}

The derivation of Eq.~\ref{fbj} for the occupation-strain interaction involved an approximation for the variation of occupation.   The full non-linear dependence on strain was retained.  This will now be relaxed applying the usual small deformation approximation.  Linearizing Eq.~\ref{mue1}  for the chemical in the strain gives
\begin{equation}
\mu = k_{\mathrm{B}} T \ln \left( \frac{c}{1 -c} \right) - I_b  - \gamma_b  + 2 \gamma_b c \left(1 -\epsilon \right)
\label{mueps}
\end{equation}
Similarly we find for the small deformation approximation to  Eq.~\ref{pstarj} 
\begin{equation}
p^\star = \gamma_b c^2 \left( 1 - 2 \epsilon \right)  - \alpha_j \epsilon
\label{pstareps}
\end{equation}
Setting $p^\star = 0$ in Eq.~\ref{pstareps} gives for the eigenstrain
\begin{equation}
\epsilon_0 = \frac{ \gamma_b c^2}{2\gamma_b c^2 + \alpha_j }
\label{eps0lin}
\end{equation}
 corresponding to a stretch 
\begin{equation}
J_0 = 1 + \epsilon_0 = \frac{3 \gamma_b c^2 + \alpha_j}{2 \gamma_b c^2 + \alpha_j}
\end{equation} 
Converting to a density using Eq.~\ref{rho2c} we find for the density at zero pessure
\begin{equation}
\frac{\rho_0}{\rho_R} = c \left( \frac{2\gamma_b c^2 + \alpha_j}{3\gamma_b c^2 + \alpha_j} \right)
\label{rho0lin}
\end{equation}
 For  a stiff lattice limit   $\alpha_j \gg \gamma_b$ and  Eq.~\ref{rho0lin} can be expanded to first order in $\kappa_j \gamma_b c^2$
 \begin{equation}
\frac{\rho_0}{\rho_R} = c \left( 1 -  \kappa_j \gamma_b c^2 \right)  
\label{rho0lin1}
\end{equation}
For repulsive interactions $\rho < c \rho_R$ as expected.  For attractive interactions

Going one step further we  ignore all dependence on the vacancy concentration except of course in the logarithm.  In this limit representing an almost ideal strained crystal Eq.~\ref{mueps} is reduced to
\begin{equation}
   \mu^0 = -   k_{\mathrm{B}} T \ln h - I_b + \gamma_b \left(1 - 2 \epsilon \right)
   \label{mueps0}
\end{equation}
This state  is hypothetical and will be referred to as a pseudo-ideal crystal.   Properties will be marked by a superscript zero. The lattice site potential in the pseudo-ideal state is still given by  Eq.~\ref{nusigma}. The pressure is found by setting $c=1$ in  Eq.~\ref{pstareps}
\begin{equation}
p^{\star0} = \gamma_b \left( 1 - 2 \epsilon \right)  - \alpha_j \epsilon
\label{pstareps0}
\end{equation}
Comparing Eqs~\ref{mueps0} and \ref{nusigma} we can say  that  the chemical potential and lattice potential play a complementary role in the thermomechanics of the crystal with a small fraction of vacancies.  The singular chemical  potential of the vacancies appears with opposite sign. In contrast, the pressure is not affected by this singularity. 
 
\section{Susceptibilities} \label{sec:response}
 
\subsection{Miscellaneous response coefficients} \label{sec:respdef}
Extending thermodynamic state space with  more variables increases the combination of possible response functions.  The example, studied in detail by LC are the elastic constants.   In the standard definition of the bulk modulus  the number of particles is constant
\begin{equation}
 B_N =  - \biggl( \frac{ \partial p^\star}{\partial \epsilon} \biggr)_{N, M}
\label{BNdef}
\end{equation}
$B_N$ is the equivalent of an elastic constant and also the number of lattice is fixed. $p^\star$ is the scaled pressure of Eq.~\ref{defpstar}.  This multiplies the bulk modulus by the same factor $v_R^c$.   We could have  made this explicit by appending a star to $B_N$ consistent with the notation for pressure.  However,   pressure,  wherever it occurs  in the derivation of response functions,  is $p^\star$ and leaving out the star except for the pressure itself should not lead to confusion. 

Alternatively,  expansion can be carried out under constant $\mu$ conditions. That this derivative makes thermodynamic sense is one of the key points of LC theory. LC define therefore  an open system bulk modulus
\begin{equation}
B_\mu =  -  \biggl( \frac{ \partial p^\star }{\partial \epsilon} \biggr)_{\mu,M} 
\label{bmudef}
\end{equation}
In the usual notation the two options are distinguished by the lower index specifying which thermodynamic variable is constrained. Note that, as for the closed system counterpart Eq.~\ref{BNdef}, the number of lattice points $M$ remains the same. Swapping pressure and strain in the derivative of Eq.~\ref{bmudef} defines an open system compressibility.
 \begin{equation}
\kappa_\mu = - \left(\frac{\partial \epsilon}{ \partial p^\star}\right)_{\mu, M}   
  \label{kappamudef}
 \end{equation}
It is understood that the number of lattice sites $M$ is always conserved. Temperature is as everywhere in this study constant.  When not specified,  these two thermodynamic constraints are assumed to be imposed in this section.

In the  extended thermodynamics of solids there are two in principle different site occupation susceptibilities.  Either  volume is kept constant
\begin{equation}
\chi_V = \left(\frac{\partial c}{\partial \mu}\right)_{V} 
\label{chiVdef}
\end{equation}
or pressure
\begin{equation}
\chi_p = \left(\frac{\partial c}{\partial \mu}\right)_{ p} 
\label{chipdef}
\end{equation}
For mixed (or cross) derivates there is a choice varying either strain,  equivalent to varying volume relative to a  fixed reference,   
\begin{equation}
\xi_V = \left(\frac{\partial c }{\partial \epsilon  }\right)_{\mu} = - \left(\frac{\partial p^\star}{\partial \mu}\right)_{V} 
\label{xiVdef}
\end{equation}
or pressure keeping the chemical potential constant. 
\begin{equation}
\xi_p = -  \left(\frac{\partial c }{\partial p^\star }\right)_{\mu} = \left(\frac{\partial \epsilon}{\partial \mu}\right)_{p} 
\label{xipdef}
\end{equation}
Second identities in Eqs.~\ref{xiVdef} and \ref{xipdef}  follow from Maxwell rules.   Eq.~\ref{xipdef} It is a formulation of the $\left( \mu, p, T \right)$ cross derivative equation
 \begin{equation} 
 \left( \frac{\partial N}{\partial p} \right)_{\mu} = - \left( \frac{\partial V}{\partial \mu} \right)_{p}
 \label{maxwell}
 \end{equation}
 Dividing by $V_R$ using Eqs. ~\ref{JV} and \ref{VRM} gives
 \begin{equation} 
 \frac{1}{v_R^c} \left( \frac{\partial c}{\partial p} \right)_{\mu} = - \left( \frac{\partial J}{\partial \mu} \right)_{p}
 \end{equation}
 Then substitution of Eqs.~\ref{vstrain} and \ref{defpstar} changes this to Eq.~\ref{xipdef}.  Eq.~\ref{xiVdef} follows from  the equivalent derivation for a $\left( \mu, V, T \right)$ system. 
 
 Measurement of the chemomechanical susceptibility $\xi_p$ requires stability under open isobaric isothermal conditions. It is a  $\left(\mu, p, T\right) $ response function which has no meaning for a liquid.  $\xi_V$ on the other hand,  describes the response of a $\left(\mu, V, T\right) $ system valid for both solids and liquids.  The difference between $ \xi_p$ and $\xi_V$ is therefore of special interest for the characterization of the distinction between the thermodynamics of liquids and solids.

 \subsection{$\left(N, V, T \right)$ bulkmodulus and eigenstrain} \label{sec:nvrep}
 In this and following subsections the response functions introduced above will be evaluated for the compressible lattice gas model of section \ref{sec:model}.  The model is intrinsically non-linear due to the $1/J$ factor in the occupation strain coupling (Eq.~\ref{fbj}).  Thermodynamic calculus requires a certain rigor to be useful.  The infinitesimal strain approximation risks obscuring fundamental thermodynamic identities.  It is therefore better to retain the full stretch $J$ as the elastic degree of freedom deferring linearization  to the final result.   Examples of LC theory for finite deformation (including shear) can be founds in Refs.~\citenum{cahn78,  mishin12,  mishin15} and \citenum{mishin20}. 
 Because $d \epsilon = d J$ for a simple volume stretch  $\epsilon$ can be directly exchanged for $J$ in the definitions of the response function of section \ref{sec:respdef}.  No modification is needed. 

The elastic response of the compressible lattice gas is derived from  the differential of the pressure Eq.~\ref{pstarj}
\begin{equation}
dp^\star =  \frac{2 \gamma_b c}{J^2} dc - \left(   \alpha_j  + \frac{2\gamma_b c^2}{J^3} \right) dJ
\label{dpstarfull}
\end{equation}
The prefactor of $dc$ will keep reappearing time and time again and deserves  its own  symbol. 
\begin{equation}
\Gamma_b\left(c, J \right) =  \frac{2 \gamma_b c}{J^2} 
\label{Gammab}
\end{equation}
The prefactor of $dJ$ is the closed system isothermal bulkmodules which will be indicated by $B_N$.  Substituting Eq.~\ref{dpstarfull} becomes less cluttered. 
\begin{equation}
dp^\star = \Gamma_b dc - B_N dJ
\label{dpstar}
\end{equation}
Next  $B_N$ is resolved in two terms
\begin{equation}
B_N = \alpha_j  + B_c
\label{BNc}
\end{equation}
$\alpha_j$ the elastic constant of the bare lattice and $B_c$ is the bulk modulus  associated with the molecular pressure $p_c^\star$ as defined in Eq.~\ref{pcstar}. Expressing $B_c$ in terms of $\Gamma_b$ of Eq.~\ref{Gammab} we have
\begin{equation}
B_c \left(c, J \right)  = \frac{ \Gamma_b c }{J} 
\label{Bc}
\end{equation}
 $B_c$ is a non-linear function of volume strain and varies with occupation taking the sign of $\gamma_b$ and can therefore  be negative. Indeed attractive interactions reduce the effective elastic constant. 

The contrast between molecular  and elastic stress is clearly manifested  in the bulk modulus  of the pseudo-ideal crystal introduced at the end of section \ref{sec:smalleps}.
\begin{equation}
B_N^0\left(\epsilon\right) = \alpha_j  + 2\gamma_b \left(1- 3 \epsilon \right)  
\label{BNeps0}
\end{equation}
In an unstrained  state ($\epsilon = 0$) the bulkmodulus is the elastic constant of the bare lattice adjusted for  molecular interaction 
\begin{equation}
B_N^0 (0) = \alpha_j + 2 \gamma_b
\label{BNeps00}
\end{equation} 
 This is as expected.   However,  while the elastic volume stress vanishes for $\epsilon = 0$,   the molecular pressure is  proportional to $\left(1 - 2 \epsilon \right)$ and remains finite (see Eq.~\ref{pstareps0}).   As explained in section \ref{sec:selfstress} an overall stress free state can only be achieved by admitting a certain amount of eigenstrain $\epsilon_0$.  Substituting  Eq.~\ref{eps0} for $\epsilon_0$ in Eq.~\ref{BNeps0} gives for the bulk modulus under zero applied pressure  
\begin{equation}
B_N^0 \left(\epsilon_0\right) = \alpha_j  + \frac{2 \gamma_b \left(\alpha_j - \gamma_b \right)}{2 \gamma_b + \alpha_j}
\end{equation}
Comparing to Eq.~\ref{BNeps00}
\begin{equation}
B_N^0(0)-  B_N^0 \left(\epsilon_0\right) =  \frac{6 \gamma_b^2}{2 \gamma_b + \alpha_j}
\end{equation}
Applying the  pressure necessary to counter the elastic eigen strain at zero external pressure hardens the system whether interactions are repulsive or attractive.

 \subsection{$\left(\mu, V, T \right)$ response functions} \label{sec:muvrep}
 The response coefficient considered next is  $\chi_V$ of Eq.~\ref{chiVdef}.  This is a  conventional grand canonical population susceptibility.   The root equation is  the differential of the chemical potential  Eq.~\ref{mue}
\begin{equation}
d\mu =\frac{k_{\mathrm{B}}T}{c \left(1 -c\right)} dc+  \frac{ 2 \gamma_b }{J}  dc  - \frac{2 \gamma_b c}{J^2} dJ 
\label{dmue}
\end{equation}
To determine the partial derivative of occupation with respect to chemical potential at constant volume
Eq.~\ref{dmue}  is first rearranged to
\begin{equation}
\left( \frac{k_{\mathrm{B}}T}{hc }   + B_c  \right) dc = d\mu   +\Gamma_b dJ
\label{dcdmudj}
\end{equation}
where we have switched over to the notation introduced in section \ref{sec:nvrep}.  Fixing volume  ($dJ =0$) we obtain for the grand-canonical susceptibility Eq.~\ref{chiVdef}
\begin{equation}
\chi_V = \frac{ \chi}{1 + \beta \Gamma_b h J}
\label{chiVhc}
\end{equation}
where $\chi$ defined as       
\begin{equation}
\chi = \beta hc
\label{chilim}
\end{equation}
 is  the "Langmuir" susceptibility. 

For qualitative understanding we can already learn from the pseudo-ideal model crystal.  The remnant of the  Langmuir susceptibility Eq.~\ref{chilim} in this limiting state is $\chi^0 = \beta h$ and Eq.~\ref{chiVhc} is simplified to
\begin{equation}
\chi_V^0 =  \beta h \left( 1 -  2 \beta \gamma_b h  \left(1 - \epsilon \right) \right)
\label{chiVhc1}
\end{equation}
 As expected, repulsive interactions reduce the uptake of particles from a reservoir with a given chemical potential.  Attractive interactions $\gamma_b < 0$ give the opposite effect.  However,  without vacancies none of this can happen.   $\chi_V$ is first order in $h$.

Continuing  with the mechanical susceptibility $\xi_V$ of  Eq.~\ref{xiVdef} we return to Eq.~\ref{dcdmudj}. Now the chemical potential is constant ($d \mu = 0$) and the differential Eq.~\ref{dcdmudj} gives an expression for $\xi_V$
\begin{equation}
\xi_V =  \frac{\Gamma_b  \chi}{1 + \beta \Gamma_b h J}
\label{xiVhc}
\end{equation}
Mechanical and chemical susceptibility  are related 
\begin{equation}
\xi_V =  \Gamma_b \chi_V
\label{xiV2chiV}
\end{equation}
The crucial difference $\xi_V$ is first order in $\gamma_b$. Without coupling to strain ($\gamma_b = 0$)  occupation is insensitive to deformation. 

\begin{figure}
\includegraphics[width=0.9\columnwidth, clip=true]{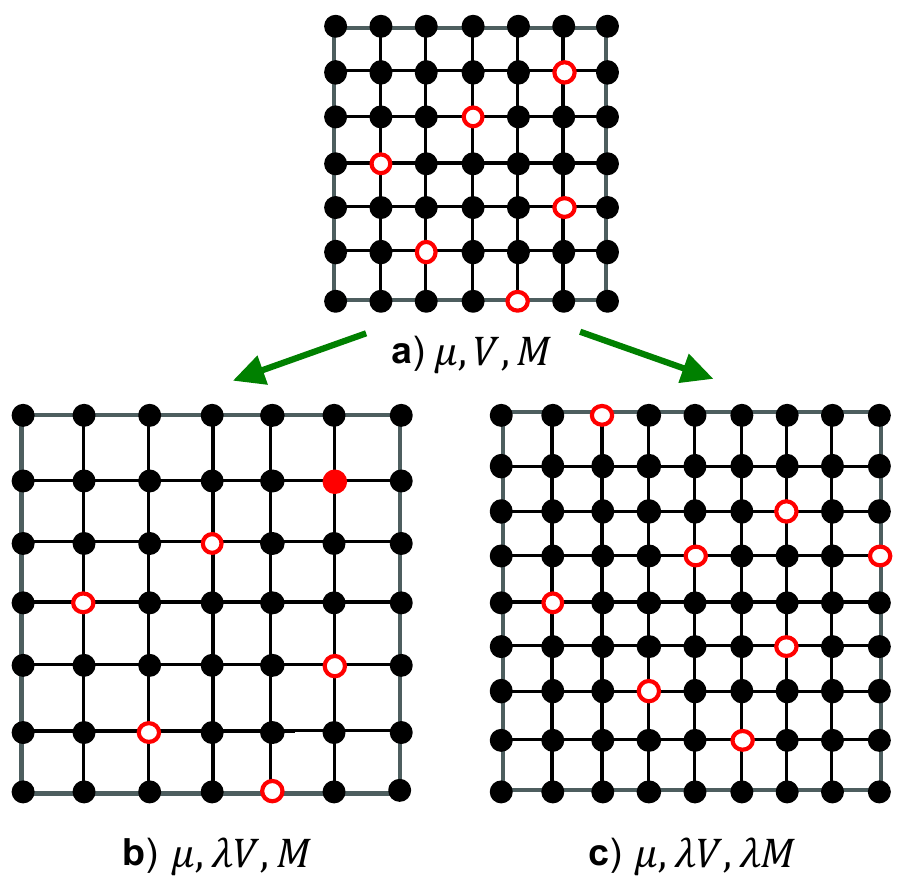}
\caption{\label{fig:openexp} Open system expansion compared to accretion.  {\bf a} shows a lattice with number of lattice sites $M$ and volume $V$ exchanging particles with a  reservoir of chemical potential $\mu$.  In {\bf b} volume has been increased by a factor $\lambda$  preserving $M$ and maintaining chemical equilibrium with the particle reservoir.   This will induce  an influx of particles but particles can only be inserted at vacant sites (solid red circle).  While increasing occupation this will effectively decrease particle density (Eq.~\ref{drhoxivj}).    In {\bf c} particle density has been restored to the initial value in {\bf a} by supplying new sites multiplying  $M$ by the same factor $\lambda$ (accretion).  For liquids there is no distinction between open system expansion and accretion. The space opened up by expansion is filled with particles with the same density as in the initial state.}. 
\end{figure}

 For more detailed analysis we again resort to the pseudo-ideal approximation
\begin{equation}
\xi_V^0 =   2 \beta \gamma_b h \left( 1 - 2\epsilon \right)
\label{xiVhc0}
\end{equation}
 For a $\gamma_b > 0$ system,  $\xi_V^0$ is positive meaning that isotropic expansion  eliminates vacancies.  Particles are moving in from the reservoir.  The increase in $N$ is limited by the lattice constraint.  In nearly complete lattices there is little room for additional  particles because the number of sites has remained the same.  This effect is  visualized in Fig.~\ref{fig:openexp}.  In the opposite case of contraction particles are pushed out to avoid the stronger repulsive interaction.  For fixed $M$  the process generates vacancies.  Now the effect of $h$ representing the vacancies already present  is harder to visualize.  However there is an additional constraint.   Chemical equilibrium must be maintained. Equating the chemical potentials in the  approximation  Eq.~\ref{mueps0}   in state 1 and 2 and exponentiating  gives for the  corresponding vacancies concentration and strain
  \begin{equation}
  h_2 = h_1 \exp \lbrack - \beta \gamma_b \left(\epsilon_2 - \epsilon_1 \right) \rbrack
 \label{hepsexp}
 \end{equation}
 Imagining 1 to be the initial state and $2$ the final state expansion ($\epsilon_2 > \epsilon_1 $)  forces  $h_2$  to be lower than $h_1$ by an amount proportional to $h_1$ consistent with Eq.~\ref{xiVhc0}  (note that the $\epsilon$ in Eq.~\ref{xiVhc0} is a secondary non-linear effect which can be  ignored). The other way around the increase in vacancy population due to contraction is also proportional  to the initial vacancy concentration 
 
 The mechanical saturation described above can be directly observed by an experiment monitoring the change in density.  The increment of density can be estimated by evaluating the differential of Eq.~\ref{rho2c}
 \begin{equation}
d \rho = \frac{dc}{v_R^c J} - \frac{c dJ}{v_R^c J^2}
\label{drhocj}
\end{equation}
Substituting Eq.~\ref{xiVdef}  for the population change this can be written as
\begin{equation}
\frac{d \rho}{\rho} =  \left( \frac{\xi_V}{c} - \frac{1}{J} \right) dJ
\label{drhoxivj}
\end{equation}
As explained,  for an almost complete crystal $\xi_V$ is small and the change in density appears as an expansion of a closed system (Fig.~\ref{fig:openexp}).  The response of liquids to increase of volume while exchanging particles with a reservoir is very different.  Any space opened up is filled up by new particles of the same density.  Crystals can be made to behave this way if the number of lattice sites is updated proportionally.  This is the accretion process discussed in section \ref{sec:accre}.  Open system expansion is therefore a signature of solid rigidity in isotropic crystals.   The pressure tensor remains throughout hydrostatic.

To derive an expression for the open system bulk modulus $B_\mu$ (Eq.~\ref{bmudef}) we start from the differential of the pressure  as given in Eq.~\ref{dpstar}  
 from which we extract the  partial derivative of pressure wrt to stretch at constant chemical potential. 
\begin{equation}
\left( \frac{\partial  p^\star}{\partial J} \right)_\mu 
 = \Gamma_b \left( \frac{\partial c}{\partial \epsilon} \right)_\mu - B_N
\label{dpstardpj}
\end{equation}
Changing sign,  the left hand side is the strain derivative defining the bulkmodulus $B_\mu$. The occupation derivative on the right hand side is  $\xi_V$ 
\begin{equation}
B_\mu = B_N -  \Gamma_b   \xi_V 
\label{bmubnxiv}
\end{equation}
$\xi_V$ is multiplied by the same factor transforming $\xi_V$ to $\chi_V$ (Eq.~\ref{xiV2chiV})  generating a second more useful expression for $B_\mu$
 \begin{equation}
B_\mu = B_N - \Gamma_b^2  \chi_V 
\label{bmubnchiv}
\end{equation}
$\chi_V$ is positive as can be seen from Eq.~\ref{chiVhc} tending to $\chi$ of Eq.~\ref{chilim} which is finite as long as there are  remaining vacancies.  

The thermomechanics implied by  Eq.~\ref{bmubnchiv} is more explicit in the pseudo-ideal crystal approximation
\begin{equation}
B_\mu^0 =  B_N^0 - 4  \beta \gamma_b^2 h \left( 1 - 2 \epsilon \right)
 \label{bmu1}
\end{equation}
 Opening up the compressible lattice  for particle exchange with the environment makes it softer.
The reduction in bulkmodulus is  quadratic in the chemomechanical coupling parameter consistent with LC theory.  However,  this effect can only be detected in crystals with a sufficient number of vacancies as was already pointed out by LC.  Note also that for zero coupling $B_\mu = B_N = \alpha_j$.   For a lattice without particle interaction it makes no difference whether the system is open or closed. Again these considerations are irrelevant if all lattice sites are occupied ($h=0$).

 \subsection{$\left(\mu, p, T \right)$ response functions} \label{sec:muprep}
 LC lattice systems are stable  under $\left(\mu, p, T \right)$ control.  This is the main theme of the paper and  this claim will now be investigated for our model system by evaluating the $\left(\mu, p, T \right)$  response coefficients defined in section \ref{sec:respdef}.  There are three such coefficients,  $\kappa_\mu$ of Eq.~\ref{kappamudef},  $\chi_p$ of Eq.~\ref{chipdef} and $\xi_p$ of Eq.~\ref{xipdef}.  The three response quantities  are related.  These relations resemble the  equations for the $\left(\mu, V , T \right)$ (grand canonical)  coefficients $B_\mu$, $\chi_p$ and $\xi_V$ studied in section \ref{sec:muvrep}.
 
We begin with deriving a connection between $\kappa_\mu$ and $\xi_p$ reusing Eq.~\ref{dpstar} for the differential of the pressure.
Rearranging gives  for the pressure derivative of  $J$
   \begin{equation}
   \left(\frac{\partial J}{ \partial p^\star}\right)_\mu = \kappa_N \left( 1 -  \Gamma_b
   \left(\frac{\partial c }{\partial p^\star }\right)_{\mu} \right)
 \end{equation}
 where $\kappa_N = 1 /B_N$ leading to the great grand canonical  inverse of the grand canonical identity Eq.~\ref{bmubnchiv}
   \begin{equation}
\kappa_\mu =   \kappa_N \left( 1 +  \Gamma_b \xi_p \right)
\label{kappamuxip}
\end{equation}
linking the compressibility $\kappa_\mu$ to  the susceptibility $\xi_p$.

 Determination of the particle susceptibility for the $\left(\mu, V, T \right)$ system in section \ref{sec:muvrep}  was based on manipulation  of the  differential of the chemical potential.  It should in principle be feasible to extend this approach to isobaric conditions but this will be much more cumbersome because of the added chemical potential dependence of the strain.   Fortunately under $\left(\mu, p, T \right)$ conditions there is an alternative route exploiting the Gibbs adsorption equations Eq.~\ref{dnudmu} and \ref{dnudp}.  The main constitutive input is now expression Eq.~\ref{nue} for the lattice chemical potential $\nu$.

 Changing over from occupation $c$ to vacancy population $h = 1 -c$ the differential of the lattice site potential is written as
 \begin{equation}
 d \nu = k_{\mathrm{B}} T \left(\frac{d h}{h} \right) - \alpha_j J dJ
 \label{dnuj}
 \end{equation}
 Both $\chi_p$ and $\xi_p$ can be determined from this differential.  
First we differentiate with respect to pressure at constant chemical potential
 \begin{equation}
 \left(\frac{\partial  \nu}{\partial p^\star} \right)_\mu =
 \frac{k_{\mathrm{B}}T}{h}  \left(\frac{ \partial  h}{\partial p^\star} \right)_\mu - \alpha_j J \left( \frac{ \partial J}{\partial p^\star} \right)_\mu 
 \label{partdnudp}
 \end{equation}
 The $\nu$ derivative can replaced by $J$ on account of Eq.~\ref{dnudp}. 
  \begin{equation}
  \left(\frac{\partial  \nu}{\partial p^\star} \right)_{\mu,T} = \frac{1}{v_R^c} \left(\frac{\partial  \nu}{\partial p} \right)_{\mu,T}  = J
\label{dnudpstar}
 \end{equation}
 The  partial derivative of $h$  is minus the partial derivative of $c$ defining $\xi_p$.  Finally, the partial derivative of $J$ is the open system compressibility $\kappa_\mu$ of Eq.~\ref{kappamudef}.  The result is 
 \begin{equation}
    \xi_p =  \beta h J \left( 1 - \alpha_j \kappa_\mu \right)
 \label{xipkappamu}
  \end{equation}
 $\xi_p$ vanishes for a system without vacancies ($h=0$).  This also means according to Eq.~\ref{kappamuxip} that $\kappa_\mu = \kappa_N$ for an ideal crystal.  If all sites are occupied  an open system becomes effectively closed.  
 Note that while the chemical potential diverges for $h \rightarrow 0$ (Eq.~\ref{mue}) the free energy and mechanical response are still defined  being equal to the response of an ideal crystal will all sites occupied. 
 
 The combination of Eqs.~\ref{kappamuxip} and  Eq.~\ref{xipkappamu} forms a coupled set of equations for $\xi_p$ and $\kappa_\mu$.   Solving for $\xi_p$ we find
   \begin{equation}
  \xi_p   =  \frac{\beta h J \left(1 - \alpha_j \kappa_N\right)} { 1  + \alpha_j \kappa_N \beta  \Gamma_b h J}
  \label{xipfull}
 \end{equation}
 Applying  various definitions and equations presented in section \ref{sec:nvrep} the numerator can be converted to a more revealing formulation
 \begin{equation}
\beta h J \left( 1 - \alpha_j \kappa_N \right) = \beta h J \kappa_N B_c = \chi \kappa_N \Gamma_b
\label{alphappa}
 \end{equation}
 To first  order approximation in $h$ the denominator of  Eq.~\ref{xipfull} can be simply ignored and we end up with 
 an encouragingly concise expression for $\xi_p$
    \begin{equation}
  \xi_p   = \chi  \kappa_N \Gamma_b 
  \label{xiph1}
 \end{equation}

 Eq.~\ref{xiph1}  for $\xi_p$  is of a similar format as Eq.~\ref{xiV2chiV} for $\xi_V$. The physical content of the result for $\xi_V$ was investigated in section \ref{sec:muvrep}.  This suggests that  comparing to $\xi_V$  might be a good option to gain some understanding of the thermodynamics of $\xi_p$.   Using Eq.~\ref{xiV2chiV} for $\xi_V$ and   Eq.~\ref{xiph1}  for $\xi_p$ we find for their ratio
\begin{equation}
\frac{\xi_p}{\xi_V}  = \frac{\kappa_N \chi}{\chi_V} = \kappa_N \left(1 + \beta \Gamma_b h J \right)
\label{xipxiv}
\end{equation}
 which in the rather brutal pseudo-ideal approximation is reduced to
  \begin{equation}
  \xi_p^0 = \kappa_N \xi_V^0
  \label{xipxiv0}
  \end{equation}
 This simple equation is one of the key results of this investigation. The implication is that absorption  induced by  application of pressure  can be understood in the same way as the change in occupation due to a change of volume.  The correspondence can be made quantitative by translating the increment  $\Delta V/ V$ in volume strain  to a change  of $-  \kappa_N \Delta p^\star$ in pressure. Manipulating the vacancy concentration  using pressure  is relatively easier in soft systems.   This sounds not unreasonable  and might even be accessible to experimental verification.  It also possibly could have been anticipated  except that  the system is free to exchange particles with the environment under constant chemical potential which makes it less obvious.  Moreover Eq.~\ref{xipxiv0} is not a general thermodynamic statement.   It relies on some rather special assumptions concerning the interactions in  compressible lattice gas model of section \ref{sec:model}.

Having determined  $\xi_p$   from the partial derivative of $\nu$ with respect to pressure at constant chemical  using Eq.~\ref{dnuj} we apply the same procedure to   find an expression for  $\chi_p$  from the partial derivative with respect to chemical potential at constant pressure
\begin{equation}
 \left(\frac{\partial  \nu}{\partial \mu} \right)_{p^\star} =
 \frac{k_{\mathrm{B}}T}{h} \left(\frac{ \partial  h}{\partial \mu} \right)_{p^\star} - \alpha_j J \left( \frac{ \partial J}{\partial \mu} \right)_{p^\star}
 \label{partdnudmu}
 \end{equation}
 The lattice site potential  derivative is replaced by minus the occupation on account of  the Gibbs-Duhem absorption isotherm Eq.~\ref{dnudmu}. The partial derivatives on the right hand side can again be related to response coefficients introduced in section \ref{sec:respdef}.   Multiplying by $\beta h$ we obtain
 \begin{equation}
  \chi_p   =  \chi -  \alpha_j  \beta h \xi_p  J 
  \label{chipxip}
 \end{equation}
 We already have an expression for $\xi_p$.  This is Eq.~\ref{xipfull}.  Inserting in Eq.~\ref{chipxip} produces a rather complicated expression which is not very informative.  The small $h$  approximation Eq.~\ref{xiph1}   leads to a  more manageable result.  It has again the form of a correction to the Langmuir susceptibility $\chi$ of Eq.~\ref{chilim}
\begin{equation}
  \chi_p  =  \chi  \left( 1 -  \alpha_j \kappa_N  \beta \Gamma_b  h J   \right)   
   \label{chilang1}
 \end{equation}
 As we did above for $\xi_p$ for a physical interpretation we will compare to the corresponding $\left(\mu,V,T\right)$  quantity. This is $\chi_V$ of 
Eq.~\ref{chiVhc} which is subtracted from Eq.~\ref{chilang1}
\begin{equation}
\chi_p - \chi_V = \chi \left( 1 +\Gamma_b \beta h J \left(1 - \alpha_j \kappa_N \right) \right)
\end{equation}
The factor $\alpha_j \kappa_N -1 $ was already dealt with in Eq.~\ref{alphappa}. Using this result $\chi_p- \chi_V$  can be reformulated as
\begin{equation}
\chi_p - \chi_V = \kappa_N \Gamma_b^2  \chi^2
\label{chipchiv}
\end{equation}
or for a pseudo ideal system
\begin{equation}
\chi_p^0 - \chi_V^0 = 4 \kappa_N \left( \gamma_b \beta h\right)^2 \left(1 - 4 \epsilon \right) 
\label{chipchiv0}
\end{equation}
The effect is quadratic in the vacancy concentration and will be very hard to detect by experiment.  The message of Eq.~\ref{chipchiv0} is that there is a great grand canonical equivalent of the grand canonical particle susceptibility at all.  Such a response coefficient is unphysical for liquids. 
 
 The story for the open system compressibility $\kappa_\mu$ starts out the same but there is surprising twist. Inserting Eq.~\ref{xiph1} in Eq.~\ref{kappamuxip} we find 
  \begin{equation}
  \kappa_\mu   =\kappa_N  \left(1 +   \chi \kappa_N  \Gamma_b^2 \right) 
  \label{kappamuh1}
 \end{equation}
 or expressed in terms of the difference with $\kappa_N$ for the pseudo-ideal lattice
  \begin{equation}
  \kappa_\mu^0  -  \kappa_N =  4 \beta h  \left(\kappa_N  \gamma_b\right)^2 \left(1 - 4 \epsilon\right)
 \label{kappamuh0}
 \end{equation}
 The change in compressibility relative to the closed system is quadratic in $\kappa_N$ and linear $h$.  For  $\chi_p^0 - \chi_V^0$ it was the other way around (Eq.~\ref{chipchiv0}),   linear in  $\kappa_N$ and quadratic in $h$.  This suggest that compressibility is the more sensitive probe of chemomechanical coupling in the agreement with the view of LC. 
 The same must hold  for the open system bulk modulus $B_\mu$ which should be formally the inverse of $\kappa_\mu$.   In fact this relation is satisfied by the expressions for $\kappa_\mu$ of Eq.~\ref{kappamuh1} and $B_\mu$ of Eq.~\ref{bmubnchiv},  which can be shown by a rather lenghty derivation which will not be given here.

\section{Microscopic connections} \label{sec:micro}

\subsection{Long range interactions} \label{sec:longrange}

The statistical mechanics of compressible Ising and lattice gas models has a long history\cite{Domb56,Essam70,Oitmaa75,Salinas87}.  The motivation was to understand the effect of elasticity on critical behavior.  It was established that coupling of the interactions between the spins to deformation gives these interactions long range character\cite{Salinas87}.   Integrated over the vibrational degrees  of freedom the effective Hamiltonian of certain simple Ising lattices resembles a so called mean field Hamiltonian originally designed to study superfluidity (the Blume-Emmery-Grifiths model)\cite{ruffo09}.  A spin in an arbitrary volume is coupled to all other spins in that volume leading to distinctly non-additive thermodynamics. 

More recently the statistical mechanics of  compressible Ising models was formulated in a unified framework which includes the long range pair interactions in the self gravitating systems of astrophysics and plasmas\cite{ruffo09,ruffo14}.   This involved the introduction of generalized ensembles adopting the approach of Hill's nanothermodynamics as already mentioned\cite{ruffo15,ruffo18}.   Of special interest  in this theory is the issue of ensemble non-equivalence and the possibility of negative heat capacity and compressibility this creates\cite{ruffo18}.  We call on this literature rooted in condensed matter physics as  theoretical basis for the generalized thermodynamics outlined in section \ref{sec:thermech}.  It should be added that thermodynamics of isothermal isobaric open systems was already discussed by Guggenheim and others without the application to real systems in mind. 

In a parallel development the contradictions created by ensemble inequivalence were also noticed in materials science.  This was instigated and guided by experimental studies of coherent decomposition in alloys\cite{cahn84,khachaturyan95,penrose99,weissmueller25}.  These systems exhibit a phase transition in the partition of their chemical components distorting but preserving the topology of the lattice. (in chemical terms no breaking of bonds).  The stress this generates is a prime example of chemomechanical coupling and a major field of application of LC theory.  Our approach is restricted to one-component systems and has not much to  contribute.  (yet) to coherent demixing in multi-component systems..   However there is a clear connection at the fundamental level of the thermodynamics of crystals which we regard as support for our interpretation of LC theory.  

\subsection{Molecular simulation} \label{sec:molsim}

The statistical mechanical theory of the critical properties of Ising lattice systems was accompanied by extensive numerical investigations.  This mini review is however limited to atomistic molecular simulation relevant to condensed matter science.  Here we should first of all mention Monte-Carlo work by David Landau in collaboration with D\"{u}nweg and others\cite{LandauDP93,LandauDP06}.  Following up on a pioneering simulation study of coherent alloy phase separation by Vandeworp and Newman\cite{NewmanKE97} Landau and coworkers set out to  carefully  quantify  various signatures of ensemble inequivalence.  This work contained repeated warnings  about the challenges posed by compressible lattice gas models and has been largely ignored possibly for this reason.  More recently research on these questions was resumed by Geissler and Dellago\cite{Dellago20,Dellago21}.  The reader is directed to these publications for an account of the state of the art of methodology and a nearly complete and up to date list of references. 

Atomistic modeling of the thermodynamics of solids was also taken up by the physical chemistry community.  A first aim was the determination of the equilibrium concentration of vacancies in simple Lennard-Jones and hard sphere crystals.   An influential work in this field is the  Monte Carlo study by Swope and Andersen (SA) \cite{swope92}. These authors are leading experts in the simulation of liquids \cite{frenkelsmit23} and adopted a theoretical perspective rather different for LC theory which was already well established at the time. The total free energy is minimized combining grand canonical (Gibbs ensemble) particle insertion methods with variation of the number of lattice sites.  The SA method  was extended by Frenkel\cite{frenkel01} and more recently by Kofke\cite{kofke18}. It is also the basis for the analysis of simulations of colloidal solids\cite{dijkstram12} and cluster crystals\cite{frenkel07}.


The number of lattice sites $M$ in the SA approach has a fundamentally different thermodynamic status compared to LC theory. $M$ of a LC crystal is the extensive state variable determining the size of the lattice. $M$ can be changed but only at the periphery such as grain and phase boundaries\cite{voorhees04,mishin15,mishin20,voorhees24}. In contrast $M$ in the SA approach is treated as an unconstrained internal degree of freedom. Formulated in terms of the generalized thermodynamics of section \ref{sec:thermech} the SA scheme amounts to enforcing a zero value for the lattice potential $\nu$.  In LC theory the value of $\nu$ is in general non-zero as explained in section \ref{sec:gdhtherm}.  Crystals with a non zero $\nu$ are considered   to be in a non-equilibrium state in the SA view. This may be true given enough time.  In fact it has been suggested that metals flow on a very long time scale\cite{sengupta18}.  However it is doubtful that the SA protocol is a suitable alternative for solving the many problems in metallurgy addressed using LC theory\cite{voorhees04}.

The open system expansion discussed in section \ref{sec:muvrep} is a good illustration of a potential discrepancy between the SA method and LC theory.  As explained,  open system expansion is a thermodynamic process consisting of isothermal increase  of volume under constant chemical potential.  For a fluid the particles influx from the environment fills up the space opened up by the expansion and particle density remains the same (accretion).  This is ultimately a consequence of the Gibbs-Duhem equation for liquids.  In solids the supply of particles is inhibited by the restriction imposed by the lattice site occupation constraint (Fig.~\ref{fig:openexp}b).  As a result particle density is reduced by open system expansion of crystals as is quantified by the $\xi_V$ coefficient.  By adjusting the number of lattice site $M$ SA interrupt open system expansion for larger than infinitesimal increments in volume changing it effectively to accretion (Fig.~\ref{fig:openexp}c).  

The vacancy formation energies computed by Frenkel\cite{frenkel01}  and Kofke\cite{kofke18}  using the SA scheme gave little reason to doubt the validity of this approach.  These results are very reasonable and must be physical.  The question remains therefore how to interpret these calculations.  In a follow up of the present paper I will argue that these results correspond to a solid in phase equilibrium with its liquid under hydrostatic stress. This imposes an additional accretion equilibrium which sets $\nu = 0$.   The proof is in the 1984 paper by Mullins\cite{mullins84}.  As is pointed out in this landmark paper  the surface thermodynamics of crystals is more complicated compared to liquid-vapour interfaces. In addition to interface energy (surface tension) solid surfaces also show elastic response.  Interface energy and surface stress are in general not equal and can even have opposite sign\cite{frenkel05,vega22} inducing a non-zero value of $\nu$ (Eq.~31a of Ref.~\citenum{mullins84}). This is another issue deserving a closer scrutiny in the SA approach.   A more detailed analysis is deferred to a forthcoming publication. 

The accretion process itself has also become  a popular subject for molecular simulation\cite{vega25}.  Monte Carlo  simulation has made major contributions (for a recent review see Ref.~\citenum{binder18}).   Also molecular dynamics (MD) simulation has entered this field\cite{frenkelsmit23}. The study of the displacement of interfaces is accessible to MD and is one of its most productive applications,  in particular for charged systems where Monte Carlo methods are less effective.  The focus of these investigations is on structure and kinetics.  Accelerated by enhanced sampling methods,  MD has led to major advances in microscopic understanding of interfaces and crystallization\cite{parrinello15a,parrinello15b,parrinello18}. 

LC theory as presented here has little to say about atomistic mechanism.  However it does raise questions relevant for the thermodynamic analysis of simulation results which often ignore elasticity.  This has become recently an issue for simulation studies of solid-liquid nucleation and interfaces,  another popular topic in MD\cite{parrinello15a,parrinello15b,parrinello18,frenkel05,michaelides16,vega20,vega22,vega25} (for reviews see Ref.~\citenum{michaelides16} and \citenum{vega25}).   It was found that surface stress is of concern for the estimation of the interface energy controlling the critical radius of a nucleus\cite{frenkel05,vega20,vega22,filion24}.  
Surface stress as distinct from surface tension is an elastic property of crystal surfaces and can be negative as was indeed observed by Vega et al in a molecular simulation of a  homogeneous crystal nucleation from the liquid phase.  The key element in our LC model is the distinction between molecular pressure and elastic volume stress.   Classical nucleation theory was originally developed for liquid gas nucleation.  Volume elasticity plays no role in a liquid droplet.  It may however of importance for the stability of a crystal nucleus.  Answering this question is a priority for future research applying the LC model proposed here.

\subsection{Classical density functional theory} \label{sec:dft}
 LC theory is based on a discrete lattice model.   The rigidity differentiating solids from liquids is built in from the start.    How can this difference arise "spontaneously"  in a first principle microscopic description?   This fundamental question is crucial for the understanding of melting and freezing. This is a huge field in condensed matter science struggling with a number of still open fundamental questions\cite{dewith23}. Here we briefly comment on some recent developments in the classical density functional theory (cDFT) of solids\cite{oettel10,fuchs15,fuchs21}. The problem addressed  in these studies is the computation of elastic constants and the formation energy of vacancies.  Vacancies are characteristic of crystals and as pointed out in earlier papers on crystal  hydrodynamics\cite{martinpc72,fleming76,fuchs10}  can be considered an order parameter in addition to shear elasticity.  This view is substantiated in the  framework of cDFT by Fuchs and colleagues\cite{fuchs15,fuchs21}.   

Volume elasticity is another  solid order parameter, related to shear elasticity but different from it.  This is the main theme of the paper.  This led us to  separate hydrostatic  pressure in a molecular and elastic component.  In a LC scheme this  is simply a constitutive assumption enforced in a model.  The question is whether there is a microscopic foundation for this separation.  Volume elasticity is a rather elusive concept in atomistic theory.  It is not a primitive atomic interaction  but an emergent phenomenon revealing itself in changes in long range statistical correlations. 

LC theory may have even deeper implications for density functional theory.  The foundation of cDFT is the statistical mechanics of the grand-canonical ensemble\cite{evans79,hansen13,schmidt22}. cDFT is in principle exact for inhomogeneous liquids  and in the course of time a number of accurate approximate functionals have been developed for hard core systems\cite{roth10}. These functionals have also been applied with some success to crystals\cite{oettel10,lutsko20}.  This suggests that at the level of exact cDFT a solid can be regarded as a liquid with spontaneously generated global inhomogeneity.   This seems to be the consensus in the DFT community\cite{Yussouff79,oxtoby81,baus84,baus87} (see however Ref.~\citenum{fuchs21}).

The question remains however whether the $\left(\mu, V, T \right)$ thermodynamic constraints as applied to  liquids are also appropriate for solids.   The message of LC theory is that volume must have an equilibrium value for a stable $\left(\mu, p, T \right)$ crystal given a vlaue of $M$.  This would be the great grand $\left(\mu, p, T\right)$ ensemble average  of volume when left to fluctuate under open system conditions.  How to determine this equilibrium volume using  the $\left(\mu, V, T\right)$  based cDFT methods? Comparison to closed $\left(N, V, T\right)$ liquids suggests that it should be possible to estimate the equilibrium volume of a crystal for fixed $\mu$ and $M$  by variational  search for a minimum in the grand potential.  
This would be consistent with a finite open system compressibility (Eq.~\ref{kappamuh1}).  If the usual Gibbs-Duhem relation for liquids remained valid the grand potential density would be insensitive to expansion and the grand potential itself would simply scale with volume lacking a stable point.   All of this is for the moment speculation based on the results of an effective Hamiltonian constructed for this purpose.  Resolving these questions at a fundamental statistical mechanical level would probably require a formal great grand canonical extension of cDFT which may prove to be identical to canonical cDFT provided sufficiently accurate density functionals are used.

\section{Conclusion} \label{sec:concl}

\subsection{Summary of non-standard features} \label{sec:summary}

The premise in this paper is that the rigidity of solids leads to violation of the Gibbs-Duhem relation in its usual  form  valid for liquids. The thermodynamic origin of the problem is that the size of the reference state for deformation is an extensive variable by itself.  If this is ignored  the elastic energy is not an extensive quantity and neither is the total energy.  Elasticity is a non-additive long range interaction. The remedy  explored in the paper is to treat  the volume of the reference state as an extensive thermodynamic state variable in addition to deformed volume and the number of particles.  This is implicit in the network principle of Larch\'{e}-Cahn (LC) theory of solids.  The number of nodes of the network (lattice sites) is postulated to be conserved under deformation similar to the number of particles. This introduces a thermodynamic force conjugate to the number of lattice sites and restores the Gibbs-Duhem relation in extended form.   The result is a form of generalized thermodynamics for solids.

The generalized thermodynamics was worked out for a single-component LC crystal.  The system is homogeneous.   Shear deformation is ignored. The volume of the reference state is uniquely determined by the number $M$ of lattice sites. $M$  is treated as  the  extensive  state variable complementing particle number $N$ and deformed volume $V$.  The thermodynamic force conjugate to $M$ was given the symbol $\nu$  in recognition of the similarity to the chemical potential $\mu$ for the particles.  The detailed thermodynamic derivations in  the paper are meant as an  illustration of how to interpret the new thermodynamic force $\nu$ and how it can be used in thermodynamic analysis.  $\nu$ plays an important role  in the driving force for accretion and can also be used to define a Gibbs absorption isotherm from which occupation and strain are determined by partial differentiation of with respect to particle chemical potential and the pressure.

A  novel constitutive element of the model  is separation of the hydrostatic pressure in a molecular component depending on the deformed density $N/V$ and a bulk elastic component determined by the volume strain proportional to $V/M$.  The deformed volume $V$ common to both terms couples the particle number to the strain.  The implication of this construction was investigated in an evaluation of the thermodynamic response functions under isothermal open isochoric $\left(\mu, V, T \right)$ and open isobaric conditions $\left(\mu, P, T \right)$.   The two susceptibilities of special interest are the change in occupation (generation of vacancies) induced by variation of the chemical potential and pressure.  It was found that the  parameter controlling the deviation from  $\left(\mu, V, T \right)$  behavior is the closed system isothermal compressibility

\subsection{Outlook} \label{sec:outlook}

To conclude we briefly discuss what is missing from our minimal model and what was left out in the thermodynamic  derivation.  What is badly missing is shear strain and stress. The isotropic constraint imposed on deformation inhibits direct  comparison to  Larch\'{e}-Cahn open system mechanical properties. Introducing shear deformation is a clear priority and is work in progress. Loosing isotropy we will have to face the well-known problem of the definition of a unique chemical potential for a solid under shear stress (commonly known as the Gibbs-prism paradox)\cite{cahn80,sprik24}.  A further limitation is that we have not yet determined the adsorption and equation of state.  The expressions for the response coefficients in section \ref{sec:response} still contain the occupation and strain as unknown quantities. This was adequate for a characterization of chemomechanical coupling  but does not allow for definite evaluation. This probably requires  numerical methods which are another item on the to-do-list. Finally, only single-phase one-component systems were considered. The revolutionary  advancement of material science  due to  Larch\'{e}-Cahn is quantitative understanding  of the behaviour of  multiphase alloys. This will also be the decisive test for the extended thermodynamics proposed here and is still wanting. 

As explained in section \ref{sec:micro} part of the motivation of this work was to bring Larch\'{e}-Cahn  theory to the attention of the physical chemistry community.  Macroscopic analysis in physical chemistry is dominated by liquid state thermodynamics, which cannot always be transferred to the solid state without adjustments.  The author hopes that this paper can be a contribution to this discussion. 

\begin{acknowledgments}
This paper is in honour of Michele Parrinello reaching another milestone in his long and spectacular career in science. His example taught me to never give up on a difficult problem. His companionship and sometimes subversive sense of humour are also much appreciated. The work reported in the present theoretical paper is part of the continuing discussion with Daan Frenkel about the thermodynamics of solids.
\end{acknowledgments}


%

\end{document}